\documentclass[preprint2]{aastex631}

\usepackage{xcolor}
\usepackage{CJKutf8}
\usepackage[normalem]{ulem}

\definecolor{myred}{HTML}{BF3465}

\def\ltsima{$\; \buildrel < \over \sim \;$}
\def\simlt{\lower.5ex\hbox{\ltsima}}
\def\gtsima{$\; \buildrel > \over \sim \;$}
\def\simgt{\lower.5ex\hbox{\gtsima}}

\newcommand {\msun}{M$_{\odot}$}

\newcommand {\Cii}{[C\textsc{ii}]}
\newcommand {\Oiii}{[O\textsc{iii}]}

\def\ltsima{$\; \buildrel < \over \sim \;$}
\def\simlt{\lower.5ex\hbox{\ltsima}}
\def\gtsima{$\; \buildrel > \over \sim \;$}
\def\simgt{\lower.5ex\hbox{\gtsima}}

\received{--}
\revised{--}
\accepted{--}
\submitjournal{ApJ}

\shorttitle{ASPIRE: [OIII] Emitters in a z=6.61 Protocluster}
\shortauthors{Champagne et al.}

\begin{document}

\title{A Quasar-Anchored Protocluster at $z=6.6$ in the ASPIRE Survey: II. An Environmental Analysis of Galaxy Properties in an Overdense Structure}

\correspondingauthor{Jackie Champagne}
\email{jbchampagne@arizona.edu}

\suppressAffiliations
\author[0000-0002-6184-9097]{Jaclyn~B.~Champagne}
\affiliation{Steward Observatory, University of Arizona, 933 N. Cherry Ave, Tucson, AZ 85721, USA}
\author[0000-0002-7633-431X]{Feige Wang}
\affiliation{Steward Observatory, University of Arizona, 933 N. Cherry Ave, Tucson, AZ 85721, USA}
\affiliation{Department of Astronomy, University of Michigan, 1085 S. University Ave., Ann Arbor, MI 48109, USA}
\author[0000-0001-5287-4242]{Jinyi Yang}
\affiliation{Steward Observatory, University of Arizona, 933 N. Cherry Ave, Tucson, AZ 85721, USA}
\affiliation{Department of Astronomy, University of Michigan, 1085 S. University Ave., Ann Arbor, MI 48109, USA}
\author[0000-0003-3310-0131]{Xiaohui Fan}
\affiliation{Steward Observatory, University of Arizona, 933 N. Cherry Ave, Tucson, AZ 85721, USA}
\author[0000-0002-7054-4332]{Joseph F. Hennawi}
\affiliation{Department of Physics, University of California, Santa Barbara, CA 93106-9530, USA}
\author[0000-0002-4622-6617]{Fengwu Sun}
\affiliation{Steward Observatory, University of Arizona, 933 N. Cherry Ave, Tucson, AZ 85721, USA}

\author[0000-0002-2931-7824]{Eduardo Ba\~nados}
\affiliation{Max Planck Institut f\"ur Astronomie, K\"onigstuhl 17, D-69117, Heidelberg, Germany}
\author[0000-0001-8582-7012]{Sarah E. I. Bosman}
\affiliation{Institute for Theoretical Physics, Heidelberg University, Philosophenweg 12, D–69120, Heidelberg, Germany}
\author{Tiago Costa}
\affiliation{School of Mathematics, Statistics and Physics, Newcastle University, Newcastle upon Tyne, NE1 7RU, UK}
\author{Melanie Habouzit}
\affiliation{Max Planck Institut f\"ur Astronomie, K\"onigstuhl 17, D-69117, Heidelberg, Germany}
\author[0000-0002-5768-738X]{Xiangyu Jin}
\affiliation{Steward Observatory, University of Arizona, 933 N. Cherry Ave, Tucson, AZ 85721, USA}
\author[0000-0003-1470-5901]{Hyunsung D. Jun}
\affiliation{Department of Physics, Northwestern College, 101 7th St SW, Orange City, IA 51041, USA}
\author[0000-0001-6251-649X]{Mingyu Li}
\affiliation{Department of Astronomy, Tsinghua University, Beijing 100084, China}
\author[0000-0003-3762-7344]{Weizhe Liu}
\affiliation{Steward Observatory, University of Arizona, 933 N. Cherry Ave, Tucson, AZ 85721, USA}
\author[0000-0002-8857-6784]{Federica Loiacono}
\affiliation{INAF - Osservatorio di Astrofisica e Scienza dello Spazio di Bologna, via Gobetti 93/3, I-40129, Bologna, Italy}
\author[0000-0001-6106-7821]{Alessandro Lupi}
\affiliation{Dipartimento di Scienza e Alta Tecnologia, Universit\`a degli Studi dell'Insubria, via Valleggio 11, I-22100, Como, Italy}
\affiliation{INFN, Sezione di Milano-Bicocca, Piazza della Scienza 3, I-20126 Milano, Italy}
\affiliation{Dipartimento di Fisica ``G. Occhialini'', Universit\`a degli Studi di Milano-Bicocca, Piazza della Scienza 3, I-20126 Milano, Italy}
\author[0000-0002-5941-5214]{Chiara Mazzucchelli}
\affiliation{Instituto de Estudios Astrof\'{\i}sicos, Facultad de Ingenier\'{\i}a y Ciencias, Universidad Diego Portales, Avenida Ejercito Libertador 441, Santiago, Chile}
\author[0000-0003-4924-5941]{Maria Pudoka}
\affiliation{Steward Observatory, University of Arizona, 933 N. Cherry Ave, Tucson, AZ 85721, USA}
\author[0000-0003-2349-9310]{Sof\'ia Rojas-Ruiz}
\affiliation{Department of Physics and Astronomy, University of California, Los Angeles, 430 Portola Plaza, Los Angeles, CA 90095, USA}
\author[0000-0003-0747-1780]{Wei Leong Tee}
\affiliation{Steward Observatory, University of Arizona,
933 N. Cherry Ave, Tucson, AZ 85721, USA}
\author{Maxime Trebitsch}
\affiliation{Kapteyn Astronomical Institute, University of Groningen, P.O Box 800, 9700 AV Groningen, The Netherlands}
\author[0000-0002-4321-3538]{Haowen Zhang \begin{CJK}{UTF8}{gbsn}(张昊文)\end{CJK}}
\author[0000-0001-5105-2837]{Ming-Yang Zhuang 
\begin{CJK}{UTF8}{gbsn}(庄明阳)\end{CJK}}
\affiliation{Department of Astronomy, University of Illinois Urbana-Champaign, Urbana, IL 61801, USA}
\author[0000-0002-3983-6484]{Siwei Zou}
\affiliation{Chinese Academy of Sciences South America Center for Astronomy, National Astronomical Observatories, CAS, Beijing 100101, China}

\begin{abstract}

We present paper II comprising a 35 arcmin$^2$ \textit{JWST}/NIRCam imaging and wide-field slitless spectroscopy mosaic centered on J0305$-$3150, a luminous quasar at $z=6.61$.
The F356W grism data reveals 124 \Oiii+H$\beta$ emitters at $5.3<z<7$, 53 of which constitute a protocluster spanning (10 cMpc)$^2$ across $6.5<z<6.8$. 
We find no evidence of any broad-line AGN in individual galaxies or stacking, reporting a median H$\beta$ FWHM of 585 $\pm$ 152\,km\,s$^{-1}$; however, the mass-excitation diagram and ``little red dot" color and compactness criteria suggest that there are a few AGN candidates on the outskirts of the protocluster.
We fit the spectral energy distributions (SEDs) of the \Oiii\, emitters with \texttt{Prospector} and \texttt{Bagpipes}, and find that
none of the SED-derived properties (stellar mass, age, or star formation rate) correlates with proximity to the quasar.
While there is no correlation between galaxy age and local galaxy density, we find modest correlations between local galaxy density with increasing stellar mass, decreasing 10-to-100 Myr star formation rate ratios and decreasing nebular line equivalent widths.
We further find that the protocluster galaxies are consistent with being more massive, older, and hosting higher star formation rates than the field sample at the 3$\sigma$ level, distributed in a filamentary structure which supports inside-out formation of the protocluster. 
There is modest evidence that galaxy evolution proceeds differently as a function of the density of local environment within protoclusters during the epoch of reionization, and the central quasar has little effect on the galaxy properties of the surrounding structure.

\end{abstract}


\section{Introduction}\label{sec:intro}
Quasars hosting supermassive black holes (SMBH) with masses exceeding $10^9$ M$_{\odot}$ are now routinely observed in the high-redshift ($z>6$) Universe \citep{Fan2023a}, but challenges still exist in identifying the mechanisms responsible for their formation and growth.
Because they reside in massive dark matter halos \citep[e.g.,][]{Costa2014a, Zhang2023d}, they are theoretically expected to trace large-scale galaxy overdensities, but observations have reached little consensus on whether this is routinely the case \citep[e.g.,][]{Kim2009a, Banados2013a, Mazzucchelli2017b, Ota2018a, Champagne2023a, Lambert2024a, Rojas-Ruiz2024a}.
While previous observations were typically limited to photometric identification of companions and/or observations in small fields of view, significant variance among quasar environments is seen even with the best observing mode available --- spectroscopy in wide fields, now available with \textit{JWST} mosaics.

Two reionization-era quasar surveys in particular, EIGER \citep{Matthee2023a, Kashino2023a, Eilers2024a} and ASPIRE \citep{Wang2023a, Yang2023a} have undertaken grism spectroscopy with \textit{JWST}/NIRCam to identify H$\beta$+\Oiii-emitters at $5.3<z<7$ in the fields of a total of 31 quasars at $6.0<z<6.9$.
Several of these fields contain notable ($\delta_{\rm gal} > 5$) overdensities of companion galaxies \citep[Wang et al. in prep,][]{Wang2023a, Matthee2023a, Champagne2024a}, but we are still lacking detailed case studies investigating these putative protoclusters.
A number of protocluster candidates at $z>5$ identified as serendipitous overdensities in survey fields have been spectroscopically confirmed with \textit{JWST} and investigated in detail \citep[e.g.,][]{Morishita2023a, Sun2023b, Helton2024a}, but targeted studies of protoclusters anchored by quasars are absent from these samples. 

To this end, this paper focuses on the quasar field J0305$-$3150, which has been exhaustively studied from optical to submm wavelengths \citep{Venemans2013a, Farina2017a, Champagne2018a, Mazzucchelli2017b, Venemans2019a, Meyer2022a, Wang2023a, Champagne2024a}.
J0305$-$3150 was originally identified in the VIKING Survey \citep{Venemans2013a, Venemans2016a} and lies at $z=6.61$, containing a $\sim10^9$\,\msun\, SMBH.
Previously known to host LBG and \Cii-emitter overdensities \citep{Ota2018a, Venemans2019a, Champagne2023a}, then \citet{Wang2023a} used Cycle 1 ASPIRE data to identify 41 galaxies at $5.4<z<6.7$ in a single NIRCam/WFSS pointing of the same field via the detection of \Oiii5007 --- given that 21 of these were within $\Delta z \pm 0.2$ from the quasar, the evidence suggested the existence of a protocluster.

Because protoclusters at $z>6$ are expected to span several tens of comoving Mpc \citep[e.g.,][]{Chiang2017a}, Paper I of this series \citep{Champagne2024a} presented Cycle 2 data consisting of a 35 arcmin$^2$ NIRCam/WFSS mosaic around J0305$-$3150 to search for additional members of the putative protocluster.
We found a total of 124 \Oiii\, emitters in the expanded mosaic, with a remarkable 53 members of the structure spanning the full NIRCam footprint ($\sim20$ Mpc) and centered on the quasar redshift.
The richness and spatial clustering of the overdensity suggests that it is a cluster progenitor similar to those seen in simulations at least out to $z\gtrsim4$ \citep[e.g.,][]{Bassini2020a, Remus2020a}, but observational definitions of ``protoclusters" at high redshift are discrepant and mostly rely on identifications of overdensities of star-forming galaxies within some redshift and projected distance cut \citep[e.g.,][]{Morishita2023a, Helton2024a, Herard-Demanche2025a}.
Much is lacking in terms of our knowledge of stellar mass assembly and galaxy/black hole evolution within the earliest protoclusters.
With spectroscopic data in hand and a suite of imaging from prior studies \citep{Ota2018a, Champagne2023a} and NIRCam \citep{Wang2023a}, we can investigate in detail the rest-optical properties of these galaxies and compare them to the general field population towards the end of the epoch of reionization (EoR).

Numerous NIRCam studies have already revealed that the properties of EoR galaxies differ substantially from lower-redshift galaxy populations.
Some of these results include systematically lower metallicities \citep{ArellanoCordova2022a}, younger stellar ages, higher equivalent width nebular emission lines \citep{Endsley2023b, Boyett2024a}, and smaller rest-optical sizes \citep{PerezGonzalez2023a}.
There has not yet been a study investigating the impact of the overdense environment on these quantities in the context of protocluster evolution.
Low-redshift clusters are dominated by massive, red elliptical galaxies \citep{Kravtsov2012a}, while protoclusters at $2<z<4$ are characterized by galaxies with high molecular gas fractions, elevated dust-obscured star formation, increased rates of mergers and AGN activity, and a reversal of the SFR-density relation \citep{Krishnan2017a, Long2020a, Hill2020a, Champagne2021a, PerezMartinez2022a, Lemaux2022a}.
We are still vitally lacking details on how the progenitors of the starbursting protocluster archetype evolve during the epoch of reionization, which we aim to pilot with the one of the first quasar-anchored protocluster studies.

Paper I of this series presented the structure of the QSO-anchored protocluster and the luminosity function of the \Oiii\, emitters, considering possible effects of the quasar on the protocluster structure and internal evolution.
In this Paper II, we perform a detailed investigation of the galaxies in a NIRCam mosaic of the field of J0305$-$3150 using a suite of Subaru, \textit{JWST} and \textit{HST} imaging and investigate their star forming properties via SED fitting and spectral analysis. 
We compare these to a field control sample of galaxies at $z<6$ from the same dataset.
We describe our dataset and reduction process with details on our catalog construction for \Oiii\, emitters and LBGs in \S\ref{sec:obs} and \S\ref{sec:catalog}. \S\ref{sec:oiii}, \S\ref{sec:sed} and \S\ref{sec:agn} discuss our SED fitting procedures and search for AGN. In \S\ref{sec:discussion} we  present our interpretation of the spatial trends within the protocluster and conclude in \S\ref{sec:conclusion}. 
Throughout this paper we assume AB magnitudes and a flat $\Lambda$CDM cosmology with H$_0$ = 70\,km\,s$^{-1}$\,Mpc$^{-1}$, $\Omega_{\Lambda}=0.7$, and $\Omega_{\rm M}$=0.3. 

\section{Observations and Data Reduction}\label{sec:obs}

\subsection{JWST Data}\label{sec:jwst}
J0305$-$3150 was observed as part of the Cycle 1 \textit{JWST} ASPIRE (A SPectroscopic survey of biased halos In the Reionization Era) program (GO \#2078, PI: F. Wang) which targets 25 $z>6.5$ quasars with F356W grism spectroscopy and F115W/F200W/F356W broadband imaging with NIRCam.
More details about the ASPIRE survey and its Cycle 2 follow-up observations can be found in \citet{Wang2023a} and \cite{Champagne2024a}, hereafter Paper I. 
Followup mosaic observations with the same setup were performed in Cycle 2 (GO \#3325, PI: F. Wang), constituting 5 additional pointings covering a total area of 35.09 arcmin$^2$.
Data reduction, background subtraction, and  astrometric alignment of the images are all described in Paper I.
That paper also describes the spectral extraction procedure that produces our final sample of 124 line emitters.
To better constrain the SEDs of our target galaxies, we also include archival imaging from \textit{HST} and Subaru.

\subsection{HST Imaging}\label{sec:hst}
We use existing \textit{HST} data in the field which was observed with ACS and WFC3 broadband filters, bracketing the Lyman break at $z\sim6-7$ (GO \#15064, PI: C. Casey).
J0305$-$3150 was observed for one orbit each with ACS F606W and F814W, and WFC3 F105W, F125W, and F160W.
Details of the data reduction can be found in \citet{Champagne2023a}, which includes the use of the standard \verb|astrodrizzle| pipeline as retrieved from MAST.
The data are resampled to a common pixel scale with the \textit{JWST} data at 0.031\arcsec/pixel using the \verb|reproject_interp| function within the python package \verb|reproject|\footnote{\url{https://reproject.readthedocs.io/en/stable/}}, which conserves flux using interpolation,  and the astrometry registered to the NIRCam F356W image.

\subsection{Subaru Data}\label{sec:subaru}
Finally, we used broadband imaging taken with Subaru Suprime-Cam as part of a program to map out the environment of J0305$-$3150 \citep{Ota2018a}.
Details of the observational setup and reduction methods can be found in \citet{Ota2018a}, who observed J0305$-$3150 for 128 minutes in $i'$, 220 minutes in $z'$, and 380 minutes in NB921 in a total usable area of 697 arcmin$^2$.
We use the reduced dataset provided by K. Ota, who in \citet{Ota2018a} matched all of the broadband data to the point spread function (PSF) of the \textit{i'} filter with a pixel scale of 0.202\arcsec/pixel.
We do not use the narrowband imaging here since the majority of our targets fall out of the redshift range for which NB921 covers Ly$\alpha$ and we do not have the sensitivity to detect continuum emission with the narrowband.
The only change we have made to this data is that the astrometry was re-registered to the F115W images. 
We opt not to PSF-match any of our other images to the Suprime-Cam data as the NB921 PSF with FWHM 0.91\arcsec\, is substantially wider than our worst spatial resolution with \textit{HST} (0.13\arcsec).

\subsection{Point Spread Function Construction}\label{sec:psf}
All of the \textit{HST} and \textit{JWST} images with PSFs smaller than that of F356W were PSF-matched to the filter with the lowest spatial resolution of our NIRCam dataset (F356W FWHM $\approx$ 0.10\arcsec).
For the three WFC3 bands and the two Suprime-Cam bands with a wider PSF, we impose an additional encircled energy correction.

We do this first by constructing empirical point spread functions for each field by stacking stars and point sources in every filter. 
We queried the \textit{Gaia} DR3 catalog and the 2MASS legacy catalog to find stars but found there were fewer than 5 non-saturated stars within the footprint of NIRCam, so we also stacked the 25 brightest point sources detected with the \verb|photutils| version of \verb|daophot|.
We spatially resample these 101$\times$101 pixel cutouts by 10$\times$ to compute any necessary sub-pixel shifts in order to recenter them.
Then we perform background subtraction, normalize them to the peak value, and perform a median stack.
Each filter was then convolved with a window function using \verb|astropy| \citep{astropy:2022}, specifically \verb|convolve_fft| to match the empirical PSF to that of F356W, and the final PSF curve-of-growth shows agreement at $<5\%$ for a 0.32\arcsec\, diameter aperture.

For the three \textit{HST}/WFC3 filters with a wider PSF than NIRCam (i.e., F105W, F125W, and F160W, all matched to a PSF FWHM $\sim0.13$\arcsec\,), we impose an additional aperture correction as follows. 
We followed the same procedure above to PSF-match F356W to F160W and ran \verb|SE++| on the native F356W detection image, using the PSF-matched F356W image as a measurement image.
We derive a correction as the ratio of flux measured in default Kron apertures in the PSF-matched image to that in the native detection image, which we apply source-by-source to the WFC3 fluxes in our master catalog.
The median total flux correction for the \textit{HST} photometry is 40\%.

For the Subaru data, the spatial resolution is coarse enough that there is significant blending of sources which are resolved by NIRCam, so the same type of aperture correction cannot be applied one-to-one for every source.
Instead, for the Subaru imaging, we run \verb|SourcExtractor++| on the Suprime Cam $z'$ image (even though the $i'$ image is deeper, our sources should drop out in $i'$); then, after applying the usual Kron aperture corrections to this separate catalog, we match sources to the NIRCam catalogs within 0.5\arcsec.
Note that we do not perform any deblending of the Subaru sources because this proved unnecessary due to the low source density of high-significance Subaru objects.

\subsection{Estimating Noise}\label{sec:zpt}
To estimate the uncertainty in our catalog fluxes, we constructed an empirical noise function as a function of aperture size since the noise is correlated and thus dependent on the number of pixels in the aperture \citep[e.g.,][]{Finkelstein2015a}. 
We measured the flux at 1000 positions distributed across each image (rejecting any apertures that contain edges, bad pixels, or real sources based on a preliminary segmentation map) with aperture sizes ranging from 1 to 25 pixels. 
A summary of all the observations including the average 5$\sigma$ depth in each image is provided in Table \ref{table:obs}.
For individual sources, the SNR is determined as the interpolated value of the noise function given the area of the extraction aperture, multiplied by the ratio between the root-mean-square (RMS) value at the pixel centroid (a measure of the local background) and the overall RMS (a measure of the global background), based on the provided weight maps where RMS = $(\rm weight\, map)^{-1/2}$.

\begin{deluxetable*}{cccccc}
\tablecaption{\label{table:obs} Details of the observations used for SED fitting.} 
\tablehead{\colhead{Instrument} & \colhead{Filter} & \colhead{FoV} & \colhead{Exposure Time} & \colhead{Average Depth} & \colhead{Matched PSF FWHM}\\ \colhead{ } & \colhead{ } & \colhead{arcmin$^2$} & \colhead{sec} & \colhead{$\mathrm{mag}$} & \colhead{\arcsec}}
\startdata
HST/ACS & F606W & 11.33 & 5760.0 & 27.1 & 0.10 \\
Subaru/HSC & i' & 697.0 & 7680.0 & 27.0 & 0.91\\
HST/ACS & F814W & 11.33 & 5760.0 & 26.7 & 0.10 \\
Subaru/HSC & z' & 697.0 & 13200.0 & 26.5  & 0.91 \\
HST/WFC3 & F105W & 4.77 & 5760.0 & 26.9 & 0.13 \\
JWST/NIRCam & F115W & 35.09 & 12282.0 & 27.2 & 0.10 \\
HST/WFC3 & F125W & 4.77 & 5760.0 & 27.2 & 0.13 \\
HST/WFC3 & F160W & 4.77 & 5760.0 & 26.8 & 0.13 \\
JWST/NIRCam & F200W & 35.09 & 12282.0 & 28.0 & 0.10 \\
JWST/NIRCam & F356W & 35.09 & 12282.0 & 28.3 & 0.10
\enddata
\end{deluxetable*}

\section{Catalog Construction}\label{sec:catalog}
\subsection{Detection Parameters}\label{sec:params}
We ran SourcExtractor++ (\verb|SE++|) \citep{Bertin2020a} in dual-image mode using F356W as the detection image with the respective weight maps for each filter.
The relevant detection parameters are the following: \verb|DETECT_THRESH| = 3.0, Kron parameters \verb|k, Rmin| = 1.2, 1.7, \verb|DETECT_MINAREA| = 10 pixels, and \verb|PHOT_APERTURES| = 12 pixels.
We used the \verb|WHT| extension as the weight maps for all measurement images.
These Kron parameters were chosen to maximize the sensitivity to faint unresolved sources, but we ran \verb|SE++| an additional time with the default Kron parameters (\verb|k, Rmin| = 2.5, 3.5) in order to derive an aperture correction for each source, defined as the ratio between the custom and default Kron fluxes in F356W and applied to the photometry in every filter.
The final photometry and signal-to-noise ratios are measured in the 0.32\arcsec\, circular apertures \verb|PHOT_APER| corrected by the ratio of the two Kron fluxes. 
We calculated the Galactic extinction in each filter by querying the \citet{Schlegel1998a} dust reddening map at the central position of each image and converting to $A_V$ using \citet{Schlafly2011a}.

\subsection{The [OIII] catalog}\label{sec:o3cat}
\begin{figure*}
\includegraphics[width=1.0\textwidth]{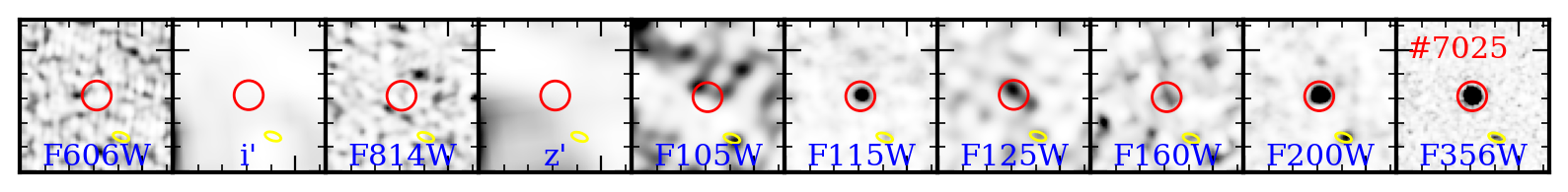}
\includegraphics[width=1.0\textwidth]{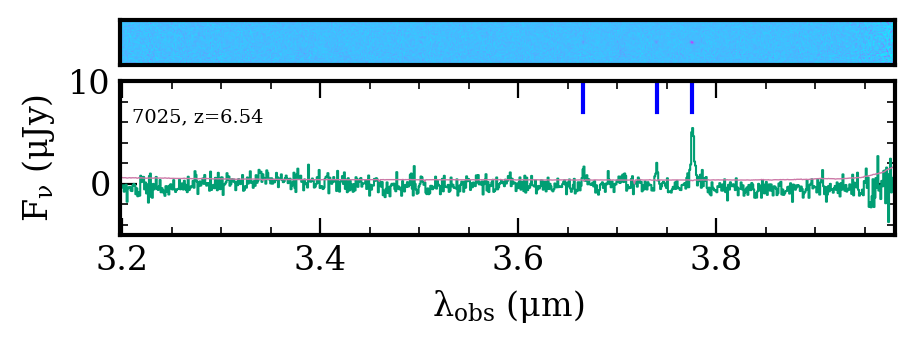}
\caption{\label{fig:example} Example cutouts and spectrum for ID7025, a member of the quasar protocluster. \textit{Top:} 2\arcsec$\times$2\arcsec cutouts in each filter from Subaru, \textit{HST} and \textit{JWST}, where the red circle denotes the circular aperture used for photometry. \textit{Middle:} 2D co-added grism spectrum highlighting the detection of three emission lines. \textit{Bottom:} Optimally extracted 1D grism spectrum (green) and error (pink) with the locations of H$\beta$, \Oiii$\lambda$4959 and \Oiii$\lambda$5007 labeled with thick blue lines.}
\end{figure*}

Details of the line-search procedure can be found in \citet{Wang2023a} and Paper I.
The 5$\sigma$ limiting flux in the deepest part of the mosaic, assuming a linewidth of 50\,\AA\, (2$\times$ the spectral resolution), is 1.2$\times$10$^{-18}$\,erg\,s$^{-1}$\,cm$^{-2}$. 
Following the extraction procedures in paper I, we here use a sample of 124 \Oiii\, emitters at $5.3<z<7$.
The protocluster at the redshift of the quasar consists of 53 galaxies at $6.5<z<6.8$ (41 of them are within $\Delta z \pm 0.05$ from the quasar, but we include 12 more galaxies that appear to be spatially associated at slightly lower and higher redshift; see Paper I for details).
We also find two serendipitous overdensities consisting of 20 galaxies at $z=5.35-5.40$ and 18 galaxies at $z=6.2-6.3$.
Finally, there are 33 ``field" galaxies not located in overdensities.
An example image and spectrum of a galaxy in the quasar protocluster is displayed in Figure \ref{fig:example}.

However, not all of these galaxies are covered by the \textit{HST} field of view, meaning these only have 3 bands of NIRCam imaging plus non-detection bands with Subaru (a full footprint of all the observations can be found in Paper I).
While we perform SED fitting for these, their properties have substantial uncertainties, and we note where necessary which galaxies' properties are not well-constrained.

Rest-frame equivalent widths are measured for the \Oiii\, emitters using the \Oiii\, line fluxes (obtained by integrating the best-fit Gaussian to the \Oiii\, doublet in the 1D spectrum) and the F356W broadband photometry with the line contribution subtracted out.
For the LBGs which are not confirmed in \Oiii\, but have robust photo-$z$ estimates (see \S\ref{sec:lbgcat}), we derive an upper limit on the EW using the 5$\sigma$ limiting line flux at the location of the LBG in the mosaic and the observed F356W magnitude.

\subsection{The LBG catalog}\label{sec:lbgcat}
We use the robust LBGs identified in Paper I. 
\citet{Champagne2023a} searched for LBGs in a wide redshift range of $\Delta z = 1.5$ due to the coarse sampling of the SED with only \textit{HST}.
Here we refit their SEDs using \texttt{EAZY} \citep{Brammer2008a} after including the NIRCam and Subaru photometry.
Of the 42 LBGs with original photometric redshifts $5.9<z<7.6$, 14 are confirmed \Oiii\, emitters in this sample, 18 no longer have $z_{\rm phot}$ consistent with the quasar redshift, and 3 are not detected in the F356W image.
In summary, we include 7 robust LBGs which have $z_{\rm phot} > 6.4$ but do not have detectable line emission.
We hypothesize that they have lower specific star formation rates which would result in undetectable \Oiii\, emission at the ASPIRE flux limit, a consequence of stochastic star formation histories (discussed in detail in Paper I).
In this study, we incorporate the 7 robust LBGs with no line identification when discussing 2D density estimates, but we do not include them in the spectroscopic analysis.

\section{SED fitting}\label{sec:oiii}

We perform two sets of spectral energy distribution (SED) fits to the photometry using \texttt{Prospector} and \texttt{Bagpipes}, in order to explore the impact of various parameter assumptions and priors on the estimates of the physical properties of interest. 
In both SED fitting runs, we impose a minimum 5\% error on the photometry, fitting all Subaru, \textit{HST}, and \textit{JWST} photometry when available, and omitting any bands if the source is not covered by that observation.

\begin{figure*}
\includegraphics[width=1.0\textwidth]{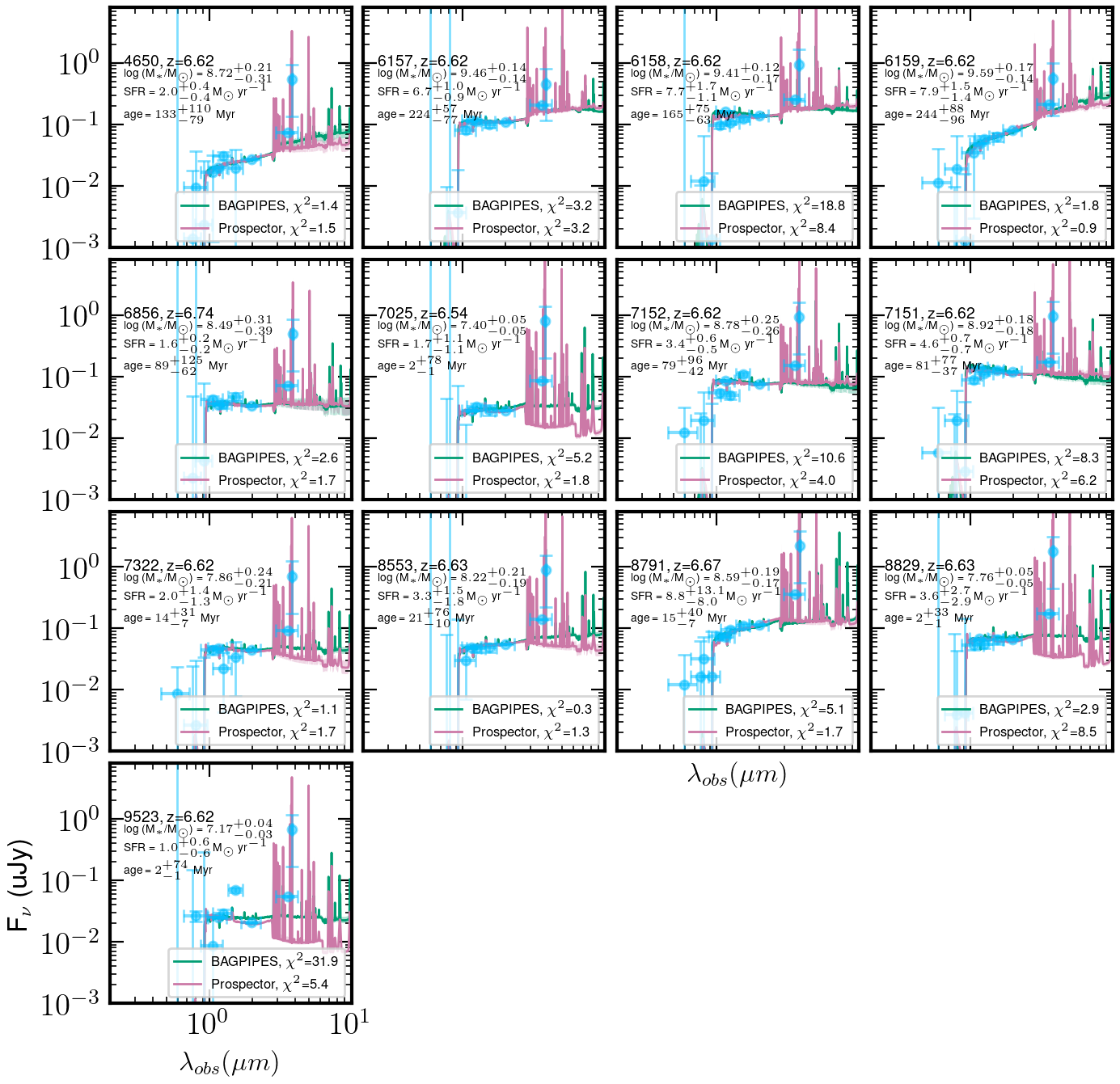}
\caption{SED fits for the spectroscopically confirmed overdensity members that are covered by \textit{HST} imaging and are located $\Delta v < 1200$\,km/s from the quasar. Best-fit \texttt{Prospector} fits are shown in pink while \texttt{Bagpipes} fits, both using a parametric delayed-$\tau$ star formation history, are shown in green, with shaded areas indicating the 16th$-$84th percentiles of the best fit spectra. The blue points are photometry from Subaru, \textit{HST} and \textit{JWST} where the $x$-errors are the width of the band and the $y$-errors are the photometric uncertainties. Inset text shows the galaxy ID, spectroscopic redshift derived from \Oiii, and best fit stellar mass, SFR, and stellar population age from our fiducial Prospector run. The $\chi^2=\Sigma(f_{obs}-f_{model})/\sigma_{obs}^2$ are shown in the legend as well.
}
\label{fig:sedmembers}
\end{figure*}

\begin{figure*}
\includegraphics[width=1.0\textwidth]{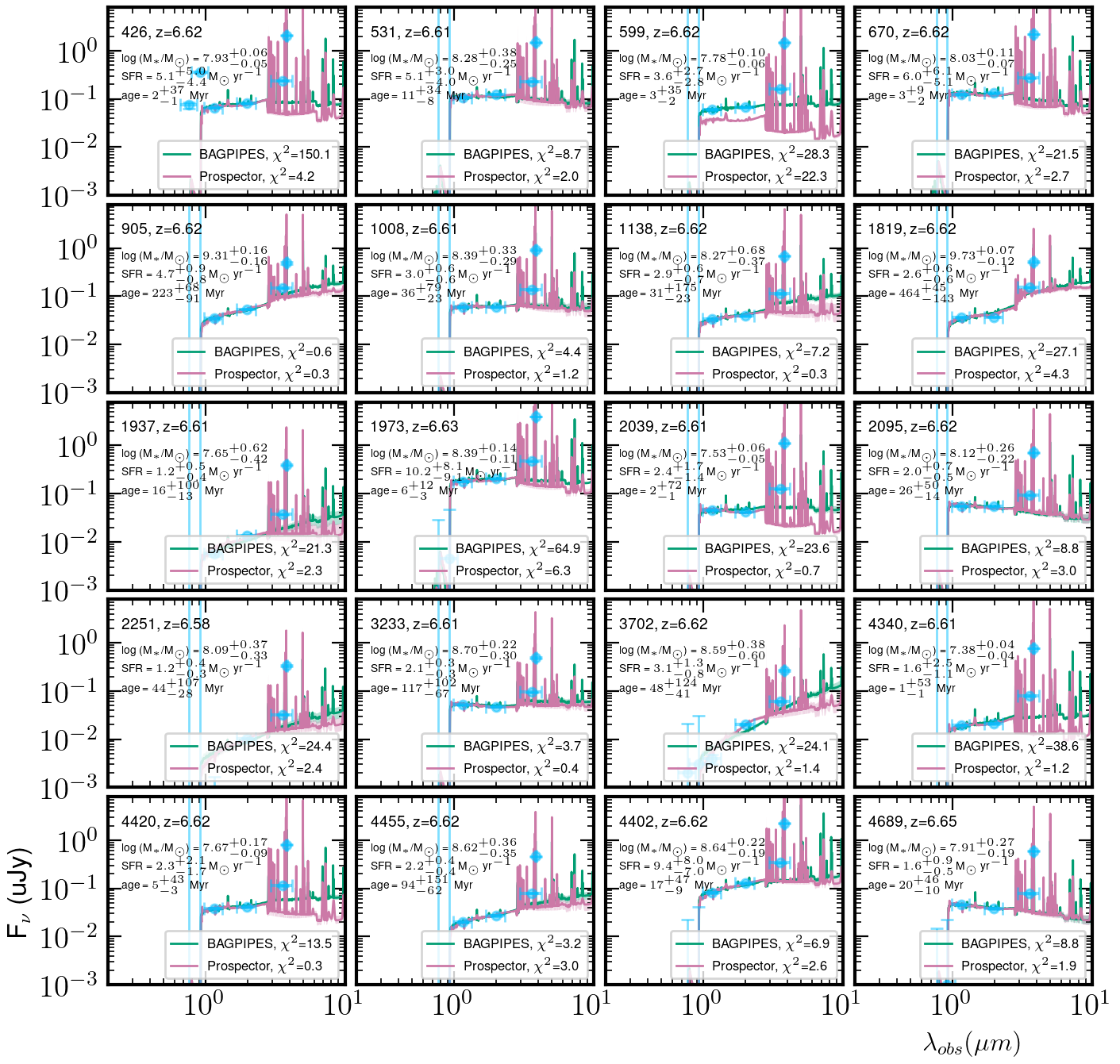}
\caption{SED fits for the spectroscopically confirmed overdensity members that are covered by \textit{HST} imaging and are located $\Delta v < 1200$\,km/s from the quasar, but do not have \textit{HST} photometric coverage. The plot symbols and colors are the same as in Figure \ref{fig:sedmembers}. The posterior distributions of the SEDs are substantially wider since there is poor coverage of the UV and optical continuum. 40/53 of the QSO overdensity members fall into this category, but they are still included in the analysis here.
}
\label{fig:sednh1}
\end{figure*}

\begin{figure*}
\includegraphics[width=1.0\textwidth]{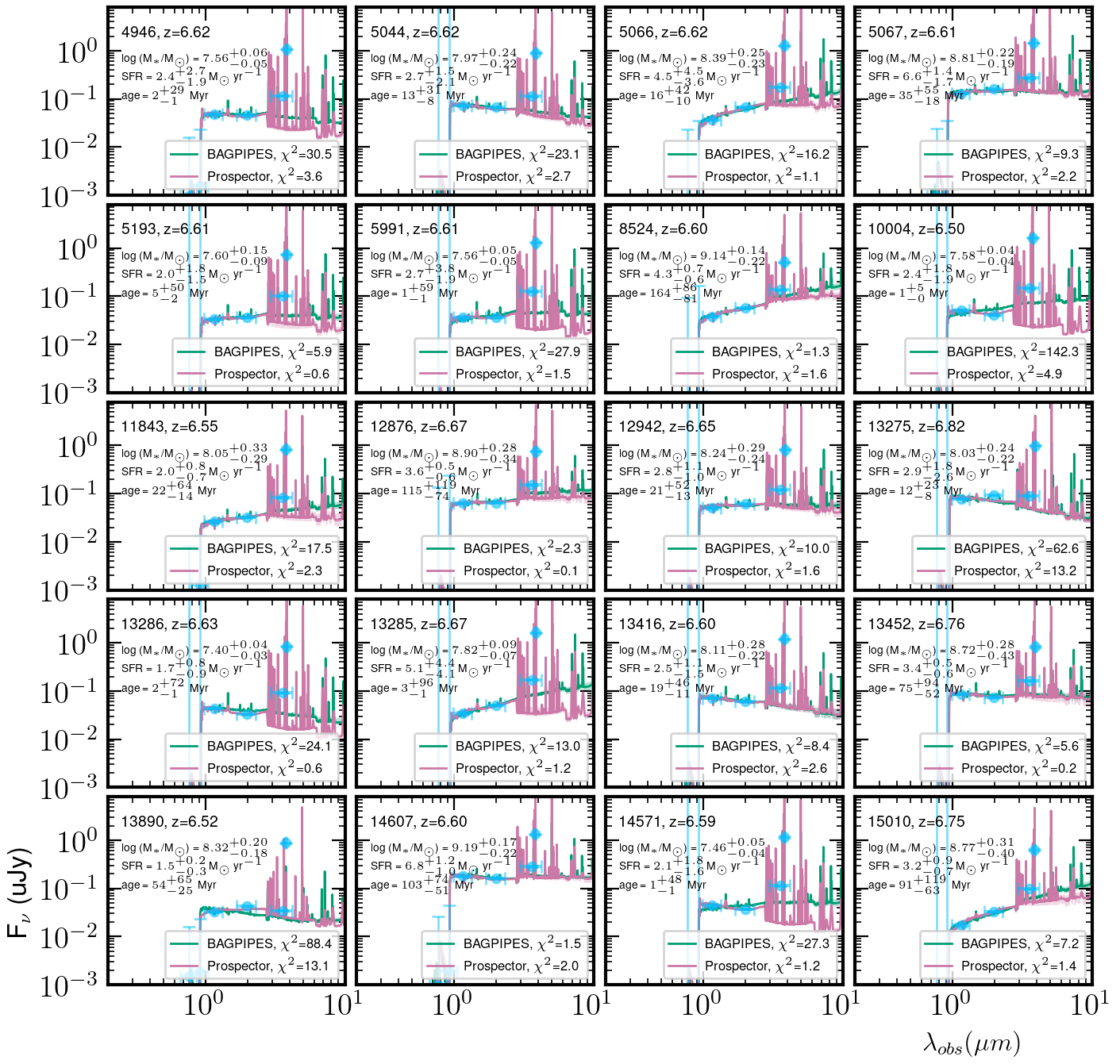}
\caption{Figure \ref{fig:sednh1} continued. 
}
\label{fig:sednh2}
\end{figure*}

We start with the use of the Bayesian Analysis of Galaxies for Physical Inference and Parameter EStimation \citep[Bagpipes;][]{Carnall2018a, Carnall2019a} SED fitting code.
\verb|Bagpipes| uses the \citet{Bruzual2003a} stellar population models, the MILES spectral library \citep{Falcon-Barroso2011a}, \verb|Cloudy| nebular emission models \citep{Ferland2017a} and a Kroupa initial mass function (IMF)\footnote{Values of stellar mass and star formation rates are converted to a Chabrier IMF for comparison with \texttt{Prospector}.}.
We use \verb|PyMultinest| \citep{Feroz2019a} to perform the sampling. 
We fix the redshift to that derived from \Oiii$\lambda$5007.
For the star formation history (SFH), we assume a delayed-$\tau$ model (i.e., SFR $\propto te^{-t/\tau})$,
allowing $\tau$ to vary within $\in [0.01, 10]$\,Gyr and $t$ to vary from 10 Myr to the age of the Universe at the \Oiii\, redshift.  
Other free parameters include the ionization parameter $U$ which is allowed to vary from log$U \in [-4, -2]$ with a flat prior, a Calzetti dust curve with a flat prior on $A_V \in [0, 3]$, and ISM metallicity (tied to the stellar metallicity) with a flat log10 prior ranging from [-2, 0.4]$Z_{\odot}$.

We also use \verb|Prospector| \citep{Johnson2019a} to explore the differences in derived properties.
\verb|Prospector| is based on the Flexible Stellar Population Synthesis (FSPS) package \citep{Conroy2009a} and uses the MIST stellar evolutionary tracks and isochrones \citep{Choi2016a,Dotter2016a}.
We use a Chabrier IMF ranging from $1-300$\,\msun\, and assume the \citet{Inoue2014a} IGM attenuation model. % with an IGM correction factor of 1.32 - 
The redshift is fixed to the \Oiii\, spectroscopic redshift as with \verb|Bagpipes|.
We assume an SMC dust curve, allowing $\tau_V \in [0.001, 5]$ with a log uniform prior; we fix the attenuation so that young stellar light is not attenuated differently from old stellar light.
The ISM metallicity is tied to the stellar metallicity, with a uniform prior ranging from
log($Z/Z_{\odot}) \in [-2.25, 0.3]$. 
We place a flat prior on the ionization parameter ranging from log$U \in [-4, -1]$.
The total mass formed has a log uniform prior from log($M_*/M_{\odot}) \in [5, 12]$ centered on log $M_*/M_{\odot}$ = 9, whose posterior is later converted to surviving stellar mass.
Following \citet{Whitler2023a}, we ignore the contribution from Ly$\alpha$, which we found has a negligible effect on the best-fit SED in our initial trial runs.
Other than Ly$\alpha$, we explicitly fit for nebular emission including the \Oiii\, doublet.

In \texttt{Prospector}, we employ the delayed-exponential (delayed-$\tau$) SFH as in \texttt{Bagpipes}, but we ran several tests experimenting with other forms of SFH priors.
We tested the commonly-used non-parametric SFH with a continuity prior, for example, but it requires too many free parameters compared to the number of data points and none of the galaxy properties are well-constrained.
A constant star formation history (CSFH) is the simplest model and is appropriate given our sparser wavelength coverage compared to e.g., JADES or CEERS, but we opted for the parameterized SFH as it is more physically motivated in the early Universe than a constant episode of star formation. 
Thus, we allow $\tau$ to vary between 0.1 and 30 with an initial guess of 1, and the age can vary between 1 Myr and the age of the Universe with a prior guess of 10 Myr, both of these quantities having log uniform priors.
The SFRs are based on the SFH posterior, averaged over 100 Myr of lookback time.

In the first iterations of our SED fitting, we noticed that both codes tended to interpret the F356W excess as a significant Balmer break rather than strong \Oiii\, emission, leading to older stellar ages, higher stellar masses, and lower star formation rates than implied by the observed strength of the \Oiii\, lines. 
Without the lever arm of additional long-wavelength imaging with the medium or broad NIRCam bands, the difference between a Balmer break and a photometric excess due to nebular line emission would be unconstrained unless there was prior knowledge of the emission line strength \citep[see, e.g.,][]{Sarrouh2024a}.
Therefore, to take advantage of the information provided by the WFSS spectra, we defined a top-hat pseudo-narrowband with a width of 100\,\AA\, centered on the \Oiii\, doublet for each galaxy and fed this to the SED codes by convolving the observed line flux with our custom narrowband (we do not include continuum as we are not sensitive to it in the WFSS observations). 
This substantially improves the agreement between the best-fit SED and the observed line amplitudes and EWs without the need for 1) modeling the variable spectral resolution of WFSS or 2) the often low-SNR continuum when using spectrophotometric inputs.

Between the two codes, we find that \texttt{Bagpipes} returns, on average, stellar masses 0.25 dex higher, SFRs 0.44 dex higher, and ages of 0.07 dex younger than that with \texttt{Prospector}; we note that there is a strong degeneracy between the age of the SED and the reported stellar mass as has been noted by other authors \citep[e.g.,][]{Endsley2021a, Whitler2023a}.
We also find that \texttt{Bagpipes} and \texttt{Prospector} tend to disagree in the shape of the rest-optical continuum due to our poor wavelength coverage beyond rest-frame 0.5\,$\mu$m; particularly, \texttt{Prospector} fits a number of large Balmer jumps leading to very young ages for the galaxies, but we cannot necessarily rule these solutions out.
In the following analysis, we use \texttt{Prospector} as our fiducial SED model as it typically returns lower $\chi^2$ values for our entire galaxy sample, and note the differences in various trends when using \texttt{Bagpipes} where relevant.
We use the median value of the posterior distributions of each galaxy property as the reported value, and the 16th and 84th percentiles as our 1$\sigma$ errors.

\section{Physical Properties of [OIII] Emitters}\label{sec:discussion2d}
We proceed with a discussion of the galaxies in the primary overdensity (referred to as ``QSO overdensity") and the field.
We will refer to the two lower-redshift overdensities (see \S\ref{sec:catalog}) as the ``control overdensities."
The remaining 33 \Oiii\, emitters at $5.4<z<6.1$ and $6.3<z<6.5$ are the ``field sample."

\subsection{SED Fitting Results}\label{sec:sed}

\begin{figure}
 \centering
  \includegraphics[width=0.8\columnwidth]{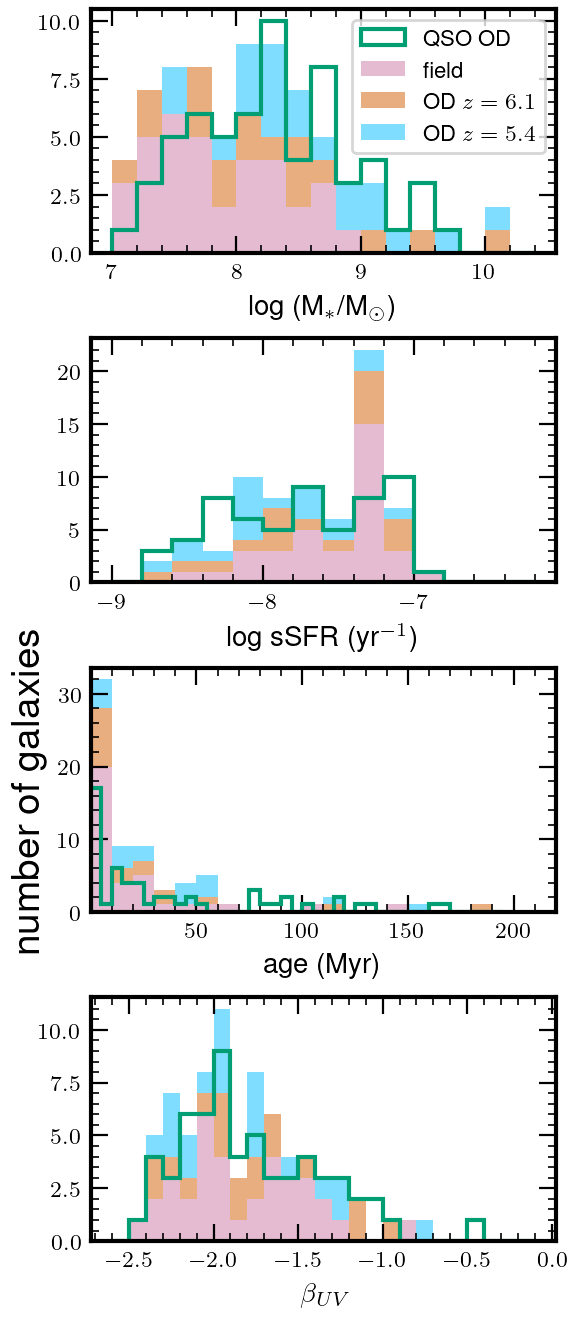}
   \caption{Histograms of SED-derived properties for galaxies in the QSO overdensity (thick green line), control overdensities, and the field sample (stacked histograms with filled colors). The QSO overdensity shows distributions of stellar mass, sSFR, stellar age, and UV slope that are overall similar to the field and control overdensities.}
    \label{fig:histos}
\end{figure}

Figure \ref{fig:sedmembers} shows the SED fits with both fitting codes for the 13 galaxies within the \textit{HST} footprint.
Figures \ref{fig:sednh1} and \ref{fig:sednh2} show the SED fits without \textit{HST} coverage, highlighting the added uncertainty in the posterior SED.
The median stellar mass of all 53 galaxies in the QSO overdensity at $z\approx6.61$ is log($M_*/M_{\odot}$) = 8.31 $\pm$ 0.08, indicating they are mostly low-mass galaxies. 
The median SFR is 4.6 $\pm$ 0.7 M$_{\odot}$\,yr$^{-1}$. 
They are young with a median stellar age of 23 $\pm$ 1 Myr, but older than the field average of 14 $\pm$ 1 Myr. 
The QSO-overdensity galaxies are fairly reddened with median UV slope $\beta=-1.81 \pm 0.06$ ($\beta \propto \lambda^{\beta}$, measured as the slope of the best-fit SED at rest-frame 1250--3000$\AA$).
This is likely due to a combination of age (ranging from 1$-$280 Myr) and moderate dust extinction ($A_V$ up to 1 mag) -- this is substantially redder than the field median of $-2.05 \pm 0.06$. 
Notably, there are a few examples within the protocluster of large Balmer jumps in low-mass galaxies with very young ages (e.g., ID8829 in Figure \ref{fig:sedmembers}), though we caution not to over-interpret these results given the sparse sampling of the rest-optical continuum.
Figure \ref{fig:histos} shows the comparisons of the two control overdensities, the QSO overdensity, and the field: overall, the distributions of SFR, age, and UV slope $\beta$ are similar.

\begin{figure}
 \centering
  \includegraphics[width=1.0\columnwidth]{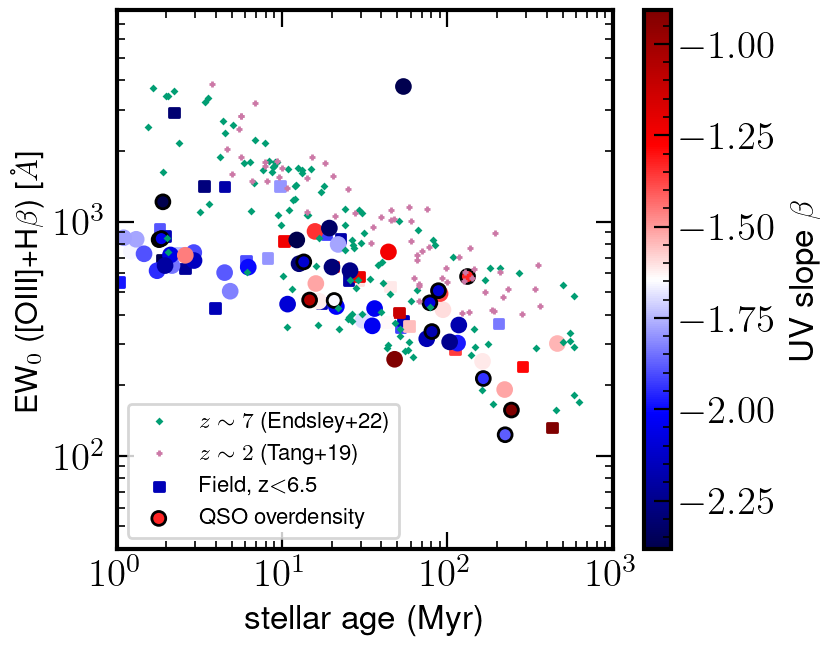}
   \caption{Rest-frame equivalent width of H$\beta$+\Oiii\, from the grism spectra versus the mass-weighted stellar age from SED fitting. The circles are \Oiii\, emitters considered to be members of the QSO overdensity, with black outlines indicating they are in the \textit{HST} footprint and have more robust SED results.
   The squares are the field sample outside of $6.5<z<6.7$ (including the control overdensities), both color-coded by the best fit UV slope $\beta$. 
   Literature field samples at $z\sim7$ \citep{Endsley2023a} and $z\sim2$ \citep{Tang2019a} are shown for comparison. Both samples from this work are consistent with the lower end of the field distribution at $z\sim7$.} 
    \label{fig:ewage}
\end{figure}

In Figure \ref{fig:ewage} we show the SED-derived stellar ages versus the rest-frame equivalent width of \Oiii+H$\beta$ as compared to the lower-$z$ ($5.4<z<6.5$) sample of \Oiii\, emitters and literature samples of photometrically-selected line emitters. 
We find that the members of the QSO overdensity are in line with the lower end of field expectations of LBGs at $z\sim7$ \citep{Endsley2023a}, though they show systematically redder UV slopes than the field-selected sample for the same stellar ages.
The median specific star formation rate (sSFR) in the QSO overdensity is $\sim19$\,Gyr$^{-1}$ while the field has median sSFR $\sim$ 23\,Gyr$^{-1}$.
This is compared to sSFR$_{\rm CSFH} =$ 18 Gyr$^{-1}$ in the ALMA-REBELS sample of $z=6-8$ LBGs which have no dust continuum detections \citep{Topping2022a}, indicating that both the field and overdensity are consistent with the average population of LBGs at this redshift.

\subsection{Occurrence Rate of AGN}\label{sec:agn}
To check whether any of our sources were AGN, we performed three tests: 1) searching for broadened H$\beta$ emission \citep[e.g.][]{Matthee2023b, Greene2024a} 2) using the mass-excitation diagram, i.e. log(\Oiii/H$\beta$) versus stellar mass \citep{Juneau2011a, Juneau2014a}, and 3) using ``little red dot" photometric criteria \citep[e.g.,][]{Greene2024a, Kokorev2024a, Akins2024a}.
However, we note that our current data is limited in discriminatory power in determining the presence of AGN.
We detect H$\beta$ in roughly half of the sample, and we use 3$\sigma$ upper limits for 65/124 galaxies.
We also do not have the wavelength coverage to access the high-ionization lines that would support the presence of AGN such as \textsc{[N iv]}$\lambda$1487, \textsc{C iv}$\lambda$1549, or He \textsc{ii}$\lambda$1640 \citep[e.g.,][]{Scholtz2023a}. 
Finally, the ``little red dot" phenomenon \citep[e.g.,][]{Matthee2023b, Harikane2023a, Greene2024a, Kokorev2024a, Akins2024a} characterized by a steeply rising red optical continuum suspected to be due to AGN, is not easily isolated without a photometric band clean of line contribution (e.g, F277W or F444W). 

\begin{figure}
    \centering
  \includegraphics[width=1.0\columnwidth]{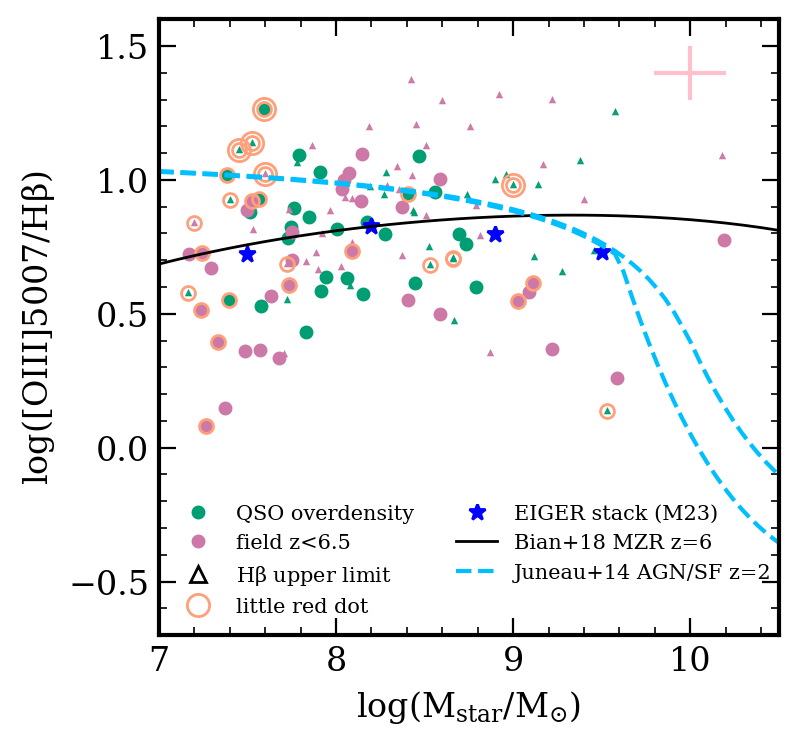}
   \caption{Mass-Excitation diagram for quasar overdensity members (green) and field members (pink, including the two other overdensities). The red bull's-eye points indicate sources that pass our little red dot criteria, with a double circle indicating that it is an AGN candidate (see \S\ref{sec:agn} for details). Typical errorbars are shown in pink in the upper right. The dashed blue lines indicate the parameterized AGN/star-forming demarcations at $z=2$ from \citet{Juneau2014a}. 
   The expected relationship using the strong-line calibration of \citet{Bian2018a} and the $z=6$ mass-metallicity relation from \citet{Ma2016a} is shown in the black line, while the EIGER \citep{Matthee2023a} stacked spectra are shown in blue.  %Despite the uncertainty in metallicity calibrators at $z=6$, 35 sources fall into the AGN regime.
   }
    \label{fig:mex}
\end{figure}

First, we fit 1D Gaussians to the H$\beta$ lines in all of the galaxies.
Of the galaxies strongly detected in H$\beta$ (approximately half of the sample), none has a FWHM exceeding 600 km/s, which is below the threshold typically considered for broad-line AGN \citep[e.g.,][]{Scholtz2023a, Larson2023a, Harikane2023a}.
Also, none of the H$\beta$ FWHMs deviate significantly from the \Oiii\, FWHM, which rules out both the presence of broad-line AGN and galactic outflows. 
We note that low-significance H$\beta$ could lead to an artificially low FWHM measurement if the broad-line signal is lost to the noise, so deeper spectroscopy is required to soundly rule out broad-line AGN.
Regardless, there is still the possibility that galaxies with point-source morphologies harbor Type 2 narrow-line/obscured AGN.

Next, we can compare the flux ratio of \Oiii/H$\beta$ to stellar mass \citep[i.e., the mass-excitation diagram or MEx;][]{Juneau2011a} to infer whether the line emission is powered primarily by star formation or an AGN.
We present this in Figure \ref{fig:mex} and compare to the demarcation between SF and AGN from \citet{Juneau2014a}. 
We find that 35/124 galaxies in the full sample are consistent with lying above the star formation boundary derived by \citet{Juneau2014a}, though we note we are comparing to an AGN/SF boundary calibrated at $z=2$ which might not hold for the typically lower-metallicity galaxies in the early Universe \citep[e.g.,][]{Shapley2023a}.
Indeed, both observational \citep[e.g.,][]{Scholtz2023a, Matthee2023a} and theoretical \citep[e.g.,][]{Hirschmann2023a} results indicate that the ratio of \Oiii\, to H$\beta$ at fixed stellar mass increases with redshift due to generally higher ionization parameters fueling higher specific SFR and, secondarily, lower metallicity within the ISM of high redshift galaxies, especially pronounced at lower stellar masses.
In Figure \ref{fig:mex} we also show the expected mass-excitation relationship based on the strong-line calibrations derived by \citet{Bian2018a} for nearby high-$z$ analogues and the mass-metallicity relationship (MZR) at $z=6$ from the FIRE simulations in \citet{Ma2016a}\footnote{The MZR relationship is scaled to the EIGER stacked observations as in \citet{Matthee2023a}.}: though there is substantial scatter, our data is consistent with the shape of the theoretically-derived MZR and the stacked observational results from EIGER at $z=6.3$ \citep{Matthee2023a}.
While 14/35 \Oiii-emitters identified in the MEx diagram above the SF boundary lie within the protocluster, we are not confident that any of these galaxies are AGN based on this selection alone.

Finally, we identify ``little red dots" \citep[LRDs; e.g.,][]{Matthee2023b, Greene2024a} which are characterized by blue UV slopes and red optical slopes \citep[only some of which show broadened Balmer lines; see][]{Harikane2023a}, suggesting contribution of an AGN to the galaxy's total light output.
To perform this search, we modify the color selection and compactness criteria from \citet{Greene2024a} to match our filter coverage.
Specifically, we impose the following color criteria\footnote{While having data in F444W would be ideal to probe the rest-optical continuum, this criteria set still searches for a red continuum slope and filters out the colors of brown dwarfs in the galaxy (which is not a concern since we only perform this search among confirmed line emitters).}:

\begin{equation}
\rm -0.5 < F115W - F200W < 1
\end{equation}

\begin{equation}
\rm F200W - F356W > 0.8
\end{equation}

\begin{figure*}
\centering
\includegraphics[width=2.0\columnwidth]{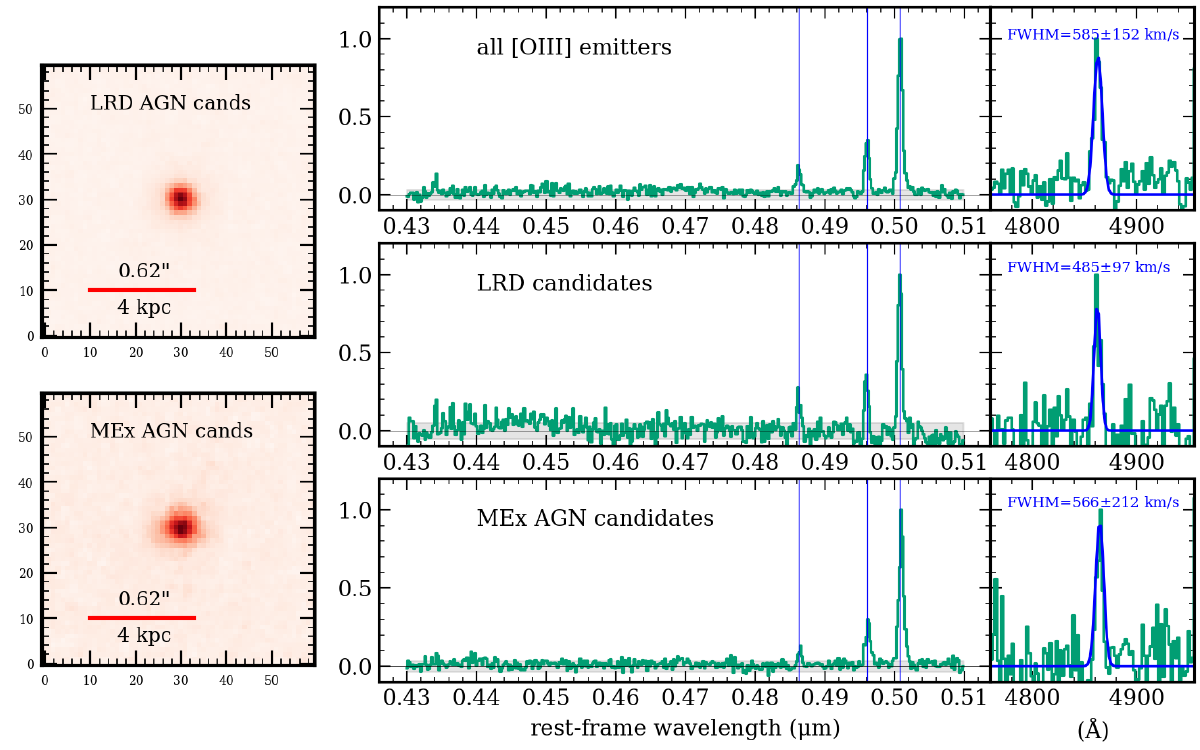}
\caption{\textit{Left:} Stacked F356W image of the little red dot \textit{(top)} and mass-excitation-selected AGN candidates \textit{(bottom)}. \textit{Middle:} Inverse variance-weighted average stack of all \Oiii\, emitters in our sample \textit{(top)}, the LRD candidates selected by color and compactness \textit{(middle)}, and AGN candidates selected by the mass-excitation diagram \textit{(bottom)}, converted to the rest frame and normalized to the peak \Oiii$\lambda$5007 flux with the typical rms in shaded grey (F$_{\nu}$ units). \textit{Right:} zoom in to H$\beta$, normalized to the peak flux, with a Gaussian fit shown in blue. %The AGN and are slightly broadened with respect to the full stack, but with high uncertainty due to a low number of sources in the stack, and thus consistent with the full stack within $1\sigma$. This broadened value is still below the formal classification for a type-1 BLR so it could hint at an obscured AGN contribution.
There is no evidence for a broadened Balmer line in any of the stacks, implying that either any contribution of AGN to the galaxy's light is obscured, or metallicity/ionization parameter effects at high redshift break down typical AGN selection criteria.
}
\label{fig:agnstack}
\end{figure*}

Lastly, we also impose a compactness criterion under the assumption that AGN-dominated galaxies would appear with a point source morphology, i.e. F(0.4\arcsec)/F(0.2\arcsec) $<$ 1.8 (performed on the F356W image, though we acknowledge that this includes line contribution). 
We find 26 sources that pass this selection, 15 of which are located within the quasar overdensity.
Five of these sources are also selected within the AGN boundary on the MEx diagram (red circles on Figure \ref{fig:mex}).
Note that these 5 sources are all low-mass galaxies (log ($M_*/M_{\odot})<$9).
If we assume the stellar masses from \texttt{Bagpipes}, we would see a median increase of 0.25 dex in stellar mass which would bring these sources closer to the AGN locus at $z=2$; still, it is interesting to note that none of the high-mass galaxies (log(M$_*$/M$_{\odot}) > 9$) above the AGN demarcation are also selected as LRDs.
This may be expected since LRDs are expected to host AGN with weak-to-moderate AGN accretion strength hosting low-mass black holes.
However, lacking both evidence for broad Balmer emission and a cleaner measurement of the optical continuum with F444W observations, we also conclude that none of these sources are \textit{a priori} AGN.

We finally perform a spectral stack to check for broadened H$\beta$ specifically among the AGN candidates selected by the LRD and MEx criteria, shown in Figure \ref{fig:agnstack}.
We also stack the full set of \Oiii\, emitters at all redshifts as a comparison sample.
Because of the variable wavelength resolution, we interpolate all the spectra onto a regularly spaced grid and de-redshift them to the rest frame before weighting them by their inverse variance and averaging them.
We find that the FWHM of H$\beta$ in all three stacks (full sample, MEx-selected, and LRD-selected) are consistent with $\sim500$\,km/s within 1$\sigma$, so we conclude that we do not yet have evidence for broad-line AGN candidates in our sample.

Due to the uncertainty in the mass-excitation demarcation for AGN and the lack of broad lines detected in the LRDs, we tentatively declare our narrow-line AGN candidates to be only the 5 sources selected by \textit{both} mass-excitation and LRD color/compactness criteria. 
Four of these lie within the quasar protocluster, while the fifth one is a field galaxy at $z=6.45$. 
This suggests that 7.5\% of the protocluster galaxies show some hint of AGN contribution, roughly in line with the 10\% contribution found for the DRC and SPT2349 protoclusters at $z\approx4$ \citep{Vito2020a, Vito2024a}.
This is lower than the findings that $\approx20$\% of JADES galaxies at $2<z<10$ show Type II AGN signatures, selected based on rest-UV and optical high-ionization lines \citep{Scholtz2023a}.
This is also lower than the $\approx$14\% of a similar parent population from JADES that  meet LRD criteria with AGN contribution based on MIR SED fitting \citep{PerezGonzalez2024a}. 
However, ours is a lower limit as our data prevent us from making firmer claims on the existence of AGN among the sample. 
Because we only see tentative evidence for AGN, we do not rerun the SED fitting codes to include any contribution from AGN \footnote{We tried to run Prospector with two additional free parameters to include the effects of an AGN torus: fractional contribution of the AGN to the bolometric luminosity $f_{\rm AGN}$ and AGN optical depth $\tau_{\rm AGN}$, but we found that this induced significant degeneracy in the stellar population fit parameters, and, further, all galaxies were consistent with 0\% contribution from the AGN given the wavelength coverage of our data.}.

\section{Discussion: A Spatial Exploration}\label{sec:discussion}
We now turn to a discussion of the general environment of the protocluster, with the goal of uncovering the effect of the overdense environment on galaxy evolution.
We first examine any influence on galaxy properties from proximity to the quasar, and then turn to an analysis of the general local density.

\subsection{The effect of the quasar on galaxy evolution}

\begin{figure*}
\centering
\includegraphics[width=1.0\textwidth]{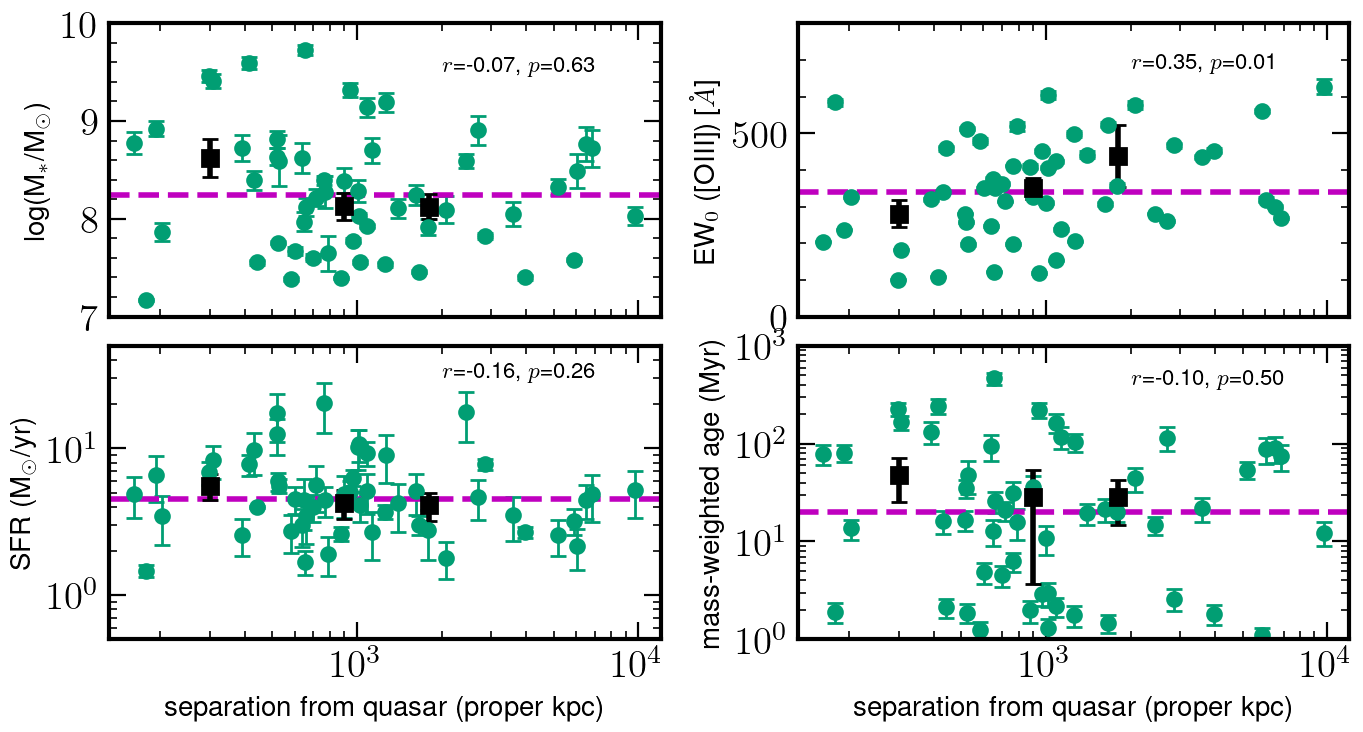}
\caption{Stellar mass, SFR, and mass-weighted age derived from SED fitting, plus equivalent width of \Oiii, as a function of radial separation from the quasar. 
The dashed pink line represents the sample median for each quantity. The black squares indicate the median values in three bins $R$ = 0--450, 450--750, and $>750$\,proper kpc. There are no strong trends between mass, age, or SFR and distance from the quasar, but there is a mild positive correlation between distance and EW (the Pearson correlation coefficient $r$ and associated $p$-value are listed in each panel).
}
\label{fig:qsoenv}
\end{figure*}

We first examine the spatial correlation of galaxy properties by their 3D positions with respect to quasar J0305$-$3150. 
In Figure \ref{fig:qsoenv} we show various SED and spectral properties with respect to the 3D distance from the quasar, incorporating the redshift offset as a proper distance. 
In general, there appears to be little to no effect on the star-forming properties of the \Oiii\, emitters directly attributable to the quasar.
The correlations between distance from the quasar and stellar mass, SFR, and age are all consistent with zero within 1 proper Mpc. 
This could hint that feedback from the quasar does not have a strong effect on galaxy evolution within its proximity zone \citep[expected to be $\sim3$\,pMpc;][]{Eilers2017a}. 
However, there is a moderate correlation between equivalent width of \Oiii\, and distance from the quasar using a Pearson correlation test ($r$-value = 0.35, $p$-value = 0.01) suggesting that the nebular line equivalent width is suppressed in the closest vicinity to the quasar. 
Also, while there is no strong correlation between stellar mass and distance from the quasar, the highest-mass galaxies with log($M_*/M_{\odot}$)$>9$ are all found closer than 1 Mpc to the quasar.
This could suggest an earlier period of rapid evolution where a large amount of stellar mass was built up in areas close to the quasar, where the recent nebular emission indicative of recent star formation ($\sim10$ Myr) is in a lull in the high-mass galaxies while star formation is ramping up at further distances.
Paper I also discusses potential feedback effects from the quasar that could be suppressing low-mass \Oiii\, emitters at close distances, where photoionization from the quasar is strongest.
Next, we will examine the effect of the dense environment while ignoring the position of the quasar.

\subsection{The general environment}
Protoclusters at high redshift are expected to be extended filamentary structures \citep{Chiang2017a}, and we have shown that our protocluster galaxies are not distributed symmetrically around the quasar. 
Therefore, we use an additional environmental measure to quantify the spatial trends of galaxy properties that eliminates the position of the quasar.
Since the galaxies lie mostly at the same redshift, we do this as a 2D projection, where we calculate the distance to the second-nearest neighbor on the sky to measure the approximate local density of sources.
Following \citet{PerezMartinez2023a} who used this measurement to quantify the environment of the famous $z=2.16$ Spiderweb protocluster, we define this density quantity $\Sigma_3$ as the following:
\begin{equation}
\Sigma_N = \frac{N}{\pi R^2_{N-1}}
\end{equation}
where $N$ = 3 and $R$ is minimum transverse radius enclosing 3 \Oiii-emitting galaxies. 
In other words, a higher value of $\Sigma_3$ indicates a shorter distance to its closest neighbors, i.e. the galaxy resides in a denser local environment.
Using this quantity, we can tease out more of the influence of the dense cosmic environment on the evolution of protocluster galaxies.
Here, we divide our full sample of \Oiii\, emitters by quasar overdensity, control overdensities, and field.

\begin{figure*}
\centering
\includegraphics[width=1.0\textwidth]{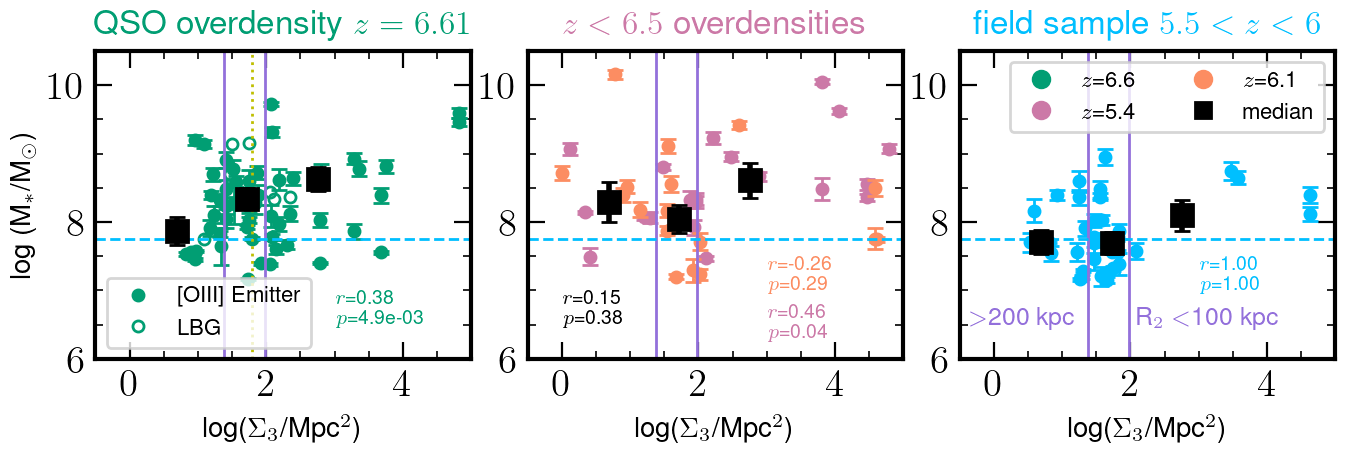}
\caption{Stellar mass derived from SED fitting as a function of local density, i.e. projected distance to the second nearest neighbor. Empty circles indicate LBGs while the filled circles are spectroscopically-confirmed \Oiii\, emitters. The left panel shows QSO overdensity members, the middle panel shows the two serendipitous ``control" overdensities, and the right panel shows the field sample. The square black points indicate the median values in three bins of local surface density, and the dotted yellow line in the left panel indicates the value of $\Sigma_3$ at the location of the quasar. All overdensitites are moderately consistent with higher median stellar masses than the field. Pearson correlation coefficient $r$ and $p$-value are shown for each sample (quasar protocluster, other overdensities, and field).
}
\label{fig:sigma3mstar}
\end{figure*}

\begin{figure*}
\centering
\includegraphics[width=1.0\textwidth]{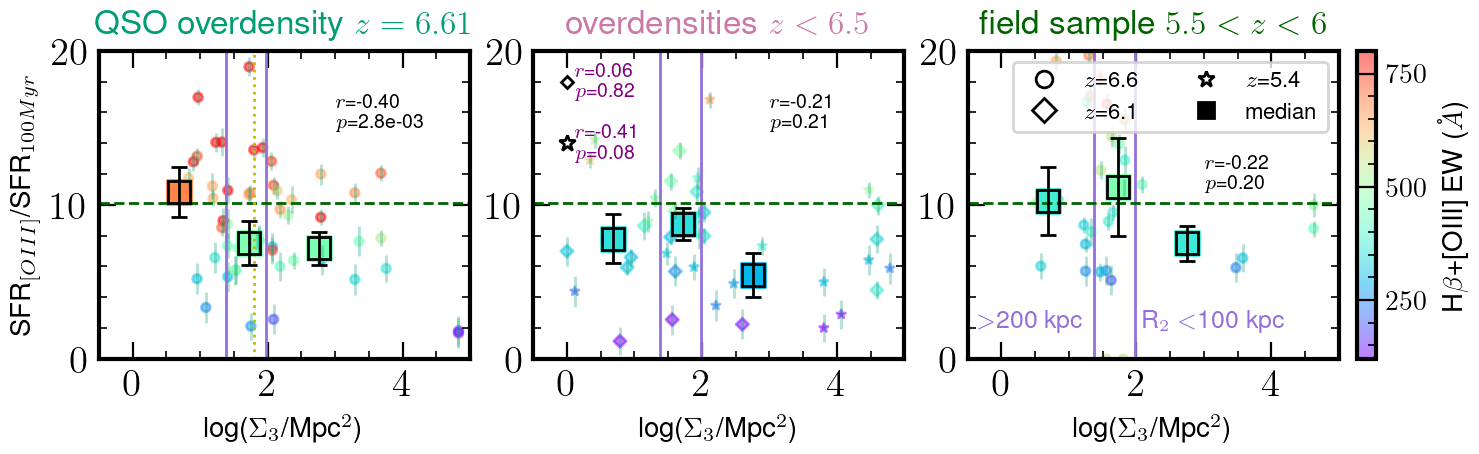}
\caption{Ratio of SFR derived from \Oiii\, luminosity (10 Myr) versus SFR derived from SED fitting (100 Myr) as a function of local density. The square black points indicate the median values in three bins of local surface density, and the dotted yellow line in the left panel indicates the value of $\Sigma_3$ at the location of the quasar. The data are color-coded by equivalent width of \Oiii+H$\beta$. In contrast to the field sample, the SFR ratio is moderately negatively correlated with density in the protocluster and weakly negative for the lower-$z$ overdensities, implying declining instantaneous star formation in high density regimes. This is corroborated by the declining EW in the highest density regime.
This could be due to a recent downturn in star formation over the last few tens of Myr, comparable to the lifetime of O stars that produce strong nebular emission \citep{Eldridge2022a}.
The median SFR ratios in all three overdensities and all three density regimes lie below the field median, implying that there was an earlier period of rapid star formation compared to the field.
}
\label{fig:sigma3ssfr}
\end{figure*}

\begin{figure*}
\centering
\includegraphics[width=1.0\textwidth]{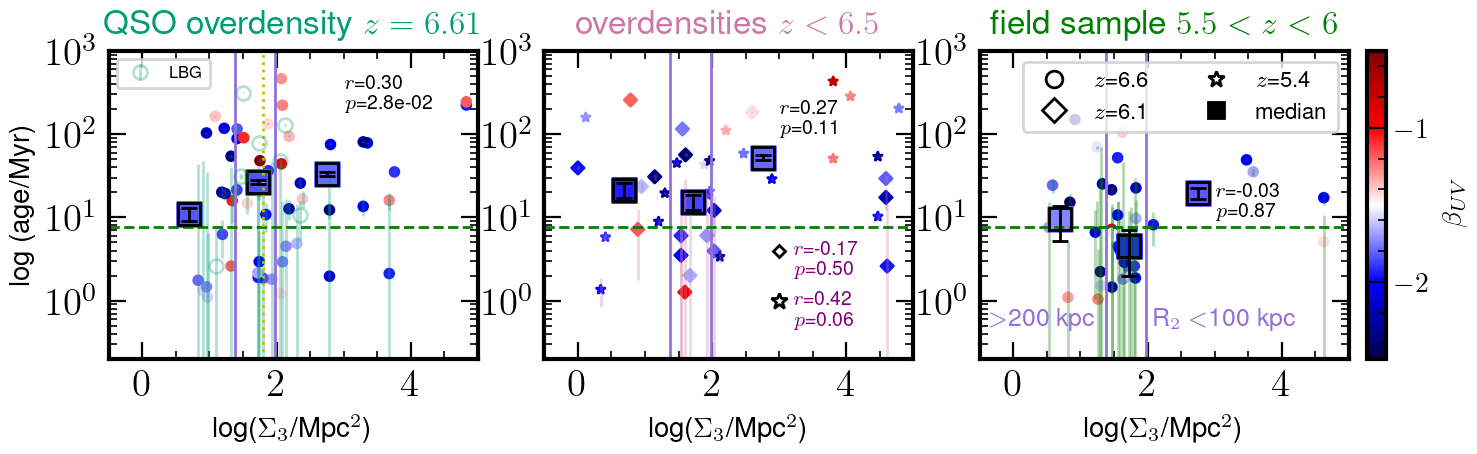}
\caption{Mass-weighted age derived from SED fitting as a function of local density with the same definition and plot symbols as the previous figure. The datapoints are color-coded by UV slope $\beta$, where we see a trend of the reddest galaxies also residing in the densest regions. In contrast to the field sample, the age is weakly positively correlated with density, implying an earlier formation time of the galaxies situated in the densest filaments of the protocluster. 
}
\label{fig:sigma3age}
\end{figure*}

Figures \ref{fig:sigma3mstar}, \ref{fig:sigma3ssfr}, and \ref{fig:sigma3age} show the spatial correlations of stellar mass, star formation rate ratios, and mass-weighted age with $\Sigma_3$ in three spatial regimes, chosen to encompass roughly equal numbers of protocluster galaxies: `dense' where $R_2$ is shorter than 100 kpc, `sparse' where $R_2$ is wider than 200 kpc, and intermediate, between 100 and 200 kpc.
In all panels of these figures, we show with large  squares the median values of each quantity within the three density bins, calculated for the all QSO overdensity members, all control overdensity members, and all non-overdensity members.
The dashed lines show the median for the whole non-overdense field sample.
Empty symbols indicate the 7 robust LBGs with $z_{\rm phot} > 6.3$, which uniformly show higher stellar masses, older ages, and lower sSFRs, in line with their lack of spectroscopic identification of nebular emission.

First, the highest stellar masses in the quasar overdensity are found in the densest regions, with a moderate but statistically significant correlation of $r=$0.37, $p$=7$\times10^{-3}$ using the Pearson test. 
The field in turn shows no such increase, while the combined lower-$z$ overdensities show a slight positive correlation but with low confidence (stronger in the $z=5.4$ overdensity; see Pearson values in Figure \ref{fig:sigma3mstar}).
Further, nearly all of the galaxies within the quasar overdensity and the two lower redshift overdensities have stellar masses above the field median, implying that they have experienced more rapid assembly of their stellar mass.
We conclude that there is a moderate relationship between the stellar mass formed and the number of close neighbors, though overall the galaxies inside overdensities are more massive than the field.

We next turn to the ratio of SFR measured on 10 Myr timescales (calculated from \Oiii) to that measured on 100 Myr timescales as a proxy for burstiness of star formation.
We measure the 10 Myr SFR using the (dust-uncorrected) SFR-$L_{\rm \Oiii}$ relationship from \citet{VillaVelez2021a} and compare these with the SFRs measured from SED fitting.
This relationship is statistically consistent with decreasing with local density within the quasar overdensity, showing $r=-0.41$ and $p$=2$\times10^{-3}$, a lower-significance anti-correlation with $r=-0.21$, $p=0.21$ in the combined lower-$z$ overdensities (again stronger in the $z=5.4$ overdensity than the one at $z=6.1$), and no correlation in the field.
In both the overdensities and the field, the \Oiii\, SFRs are higher than the 100 Myr SFR measured from the SED, but this is an expected selection effect since our survey was designed to detect bright \Oiii\, emitters.
However, in all of the overdensities, the median SFR ratio lies below that of the field median, implying that the instantaneous SFR has fallen faster in denser regions.
Together with the moderate positive correlation between local density and stellar mass, we interpret this as mild evidence for an earlier period of rapid mass assembly in denser parts of protoclusters or overdensities.
Still, we repeat that the SFR ratios are all $>1$, indicating continuous ongoing star formation in all regions of the overdensities, with none of them qualifying as quiescent (log sSFR $< - 9.5$).

Figure \ref{fig:sigma3ssfr} is further color-coded by \Oiii+H$\beta$ EW, which should be correlated with sSFR on 10 Myr timescales \citep[e.g.,][]{Wilkins2023a}.
We indeed find the lowest nebular line equivalent widths in the galaxies with the lowest SFR ratios: %which are more numerous in the denser regimes in all three overdensities: 
the median EW decreases from $\sim$800\AA\, to $\sim$470\AA\, from the lowest to the highest density bin for the quasar overdensity.
We conclude that the filamentary structures in the quasar protocluster have a diversity of star-forming states --- it hosts a large population of highly star-forming galaxies with strong nebular emission alongside galaxies with weaker line emission and lower star formation rates.
In the densest areas of the protocluster, the galaxies have likely experienced a recent downturn in their star formation while it ramps up in the outer parts of the filaments, as demonstrated by the moderate negative correlation with local density.

Finally, Figure \ref{fig:sigma3age} shows the relationship between local density and age, color-coded by UV slope $\beta$.
The median ages in the three density bins of the quasar protocluster are once again consistent with a moderate correlation ($r=0.34$, $p=0.01$), though we see significant scatter as in the SFR ratios: a dual population of extremely young blue galaxies alongside older, redder galaxies. 
There is also a weak positive correlation in the lower-redshift overdensities; however the two structures appear to have different individual distributions, with age strongly positively correlating in the $z=5.4$ overdensity but not in the $z=6.1$ overdensity.
The field is consistent with there being no correlation between density and age, and the galaxies are uniformly fairly blue.
The average slope $\beta$ increases from a median of $-2.0$ to $-1.7$ across the density bins in the quasar protocluster, so the reddening is likely due to a combination of both age and dust given the scatter.
Overall, the three overdensities show weak evidence for having formed galaxies earlier in the denser regimes, but it could also be consistent with a random distribution of ages.

\subsection{Overall Trends}
We have shown that on average, all three overdensities (the QSO overdensity and two lower-redshift overdensities) show similar spatial trends, while there is no correlation between galaxy properties and distance from the quasar.
Taken together, we can conclude that it is the density of environment in a general sense that affects galaxy evolution rather than being influenced directly by a luminous quasar.
The strongest spatial correlation is that the most massive galaxies are located in the densest regimes, followed by a moderate correlation between star formation burstiness and local density and finally a weak positive trend between age and local density.

\begin{figure*}
 \centering
  \includegraphics[width=2.0\columnwidth]{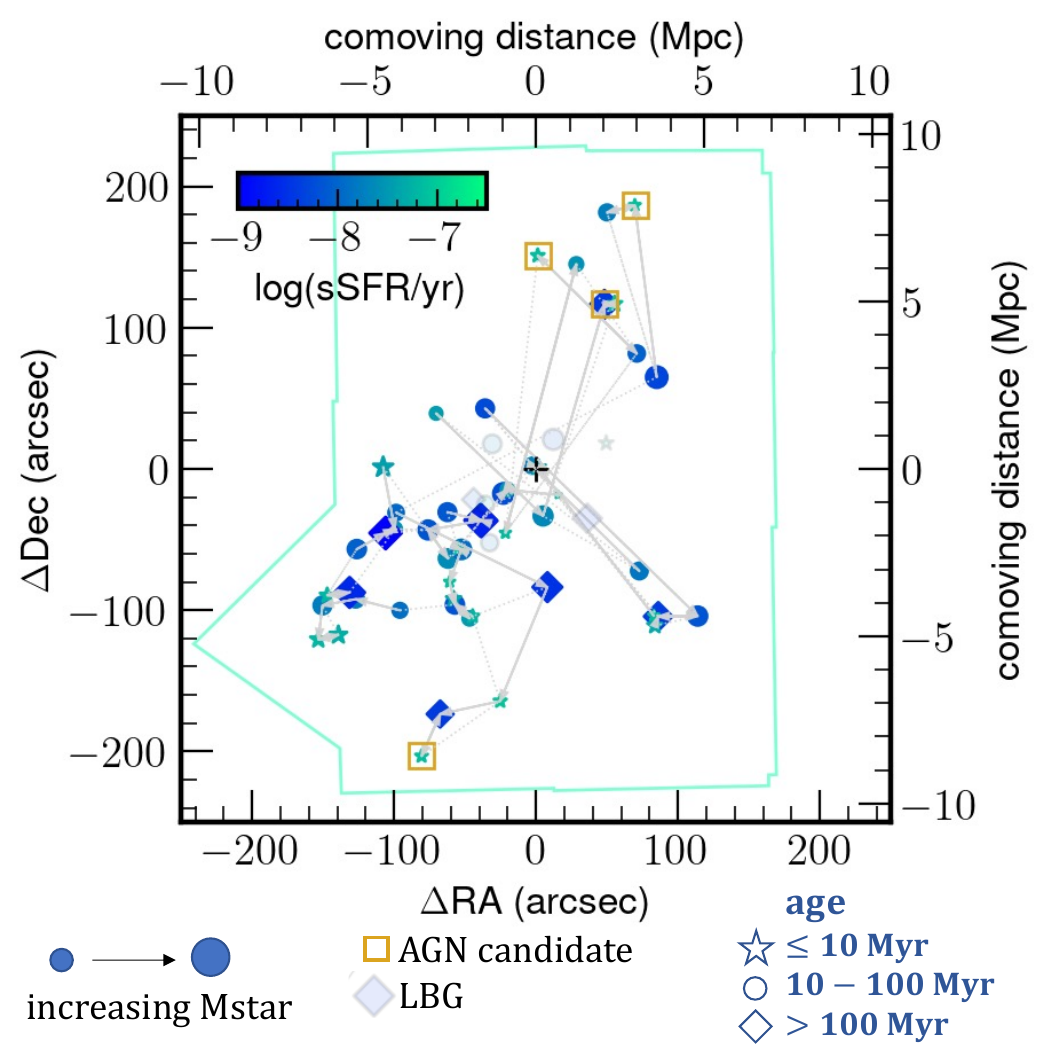}
   \caption{Schematic of the protocluster surrounding the quasar, indicated by the black cross at (0,0). 
   The solid grey arrows connect a galaxy to its closest neighbor in 3D space, and the dotted grey lines connect the second closest neighbors.
   The light green line indicates the footprint of the mosaic. The relative size of the points scales with stellar mass, and the colorbar indicates log(sSFR). The different symbols correspond to three age bins of $<$10 Myr (stars), 10--100 Myr (circles) and $>$100 Myr (diamonds). The faded galaxies are the robust LBGs still included in this sample. The golden squares indicate AGN candidates selected by both LRD and MEx criteria. We see that the largest galaxies are also the oldest and have the lowest sSFR, and are spread out along the filament extending away from the quasar. These galaxies are surrounded by more numerous young, more highly star-forming galaxies in their immediate vicinity.}
    \label{fig:3d}
\end{figure*}

There is clear filamentary structure extending at least 12 cMpc (see Paper I and Figure \ref{fig:3d}), the protocluster galaxies are generally fairly randomly distributed across the sky in a wide range of local neighbor density.
From the correlations between SED-derived properties and local density, it is clear that there is some evidence for a transition in galaxy properties on nearest-neighbor distance scales of hundreds of ckpc.
There are only moderate correlations due to significant scatter in all three distributions of mass, SFR ratios, and age.
However, a KS test shows that the distributions of stellar mass, age, and SFR are drawn from different samples between the protocluster and field at a confidence level $p<0.05$: in the protocluster, the median stellar masses are higher by $\sim$0.35 dex, the average stellar age is older by $\sim$0.3 dex, and the average sSFR is lower than the field median by $\sim$0.2 dex, all consistent with the protocluster having experienced preferentially faster mass assembly on timescales 10$-$100 Myr prior to observation.

We conclude that a rising tide raises all boats: the protocluster as a whole experiences accelerated evolution rather than such evolution being hyper-localized, though the denser areas appear to form faster and slightly earlier.
One might expect spatial trends to be weak at this stage of protocluster formation, as the density contrast with respect to the field is expected to be lower and the total spatial extent wider than that of protoclusters in more evolved states at $z<6$ \citep{Chiang2017a}.
Over time, one might expect the increasingly dense environment to have an effect on the star formation activity in an inside-out fashion.
This is readily apparent in the local density figures: the overdensity at $z=5.4$ is much denser spatially and shows much stronger departures from the field averages in sSFR and stellar age.
\citet{Forrest2024a} observes a weak but positive correlation between stellar mass and overdensity at $z\sim3.5$ and suggests pre-processing of galaxies wherein they build up the majority of their stellar mass before cluster virialization, consistent with our findings at $z=6.6$.

We finally visualize the protocluster environment in a 3D plot in Figure \ref{fig:3d}.
The quasar is placed at the origin and all angular separations are converted to comoving Mpc.
The size of the symbol increases with stellar mass, the colorbar indicates sSFR increasing from blue to green, and the symbol type indicates age ranges of 0$-$10 Myr (stars), 10$-$100 Myr (circles), and $>$100 Myr (diamonds).
The grey lines connect each galaxy to its closest neighbor in 3D space, which immediately makes the NW-SW filament apparent.
We see that the oldest and highest-mass galaxies are found in the filament extending away from the quasar, which in turn have the lowest sSFRs (large blue diamonds): star formation is instead enhanced in their smaller companion galaxies surrounding the most massive galaxies. 
In Paper I we suggested that a dearth of \Oiii\, emitters at the faint end of the luminosity function was due to temporary scattering of galaxies' nebular emission below the grism detection limit.
Taken together, these relationships imply that galaxies within the protocluster undergo stochastic periods of star formation which was enhanced in the densest areas of the filaments in an epoch a few Myr prior to observation.

Finally, we also find that all the AGN candidates selected by both mass-excitation and LRD criteria are on the outskirts of the protocluster, implying a simultaneous event triggering AGN activity.
This could be due to gas inflows along the cosmic web that are feeding the lower-mass galaxies on the outskirts of the overdensity.

Since the NIRCam coverage is deepest in the modules centered on the quasar, we should expect more faint galaxies in the center of the mosaic than in the outskirts. 
However, the \Oiii-emitter flux distribution is similar in all areas of the map. 
Conversely, bright and massive galaxies would have been easily detectable in the center, yet the only massive galaxies (log($M_*/M_{\odot}) > 9$) within 2 cMpc are LBGs (i.e., non-\Oiii-emitters).
Given that the \Oiii\, LF in the quasar overdensity is biased to faint luminosities (Paper I), we note the selection effect of an \Oiii\, survey selecting only for highly star-forming galaxies, potentially missing more massive galaxies with lower sSFR beyond our \textit{HST} footprint.

In the quasar overdensity population, the overall higher stellar masses and mass-weighted ages above the field suggests earlier stellar mass buildup along filamentary nodes, while the decreasing sSFR and SFR ratios suggest that some of the more massive galaxies are approaching a `lull' phase in their star formation.
We argue that this is the case because of the accompanying decreasing of nebular line equivalent widths with increasing spatial density, indicating that the declining specific star formation has been in effect for longer than the lifetime of O stars, though low metallicity or high ($>$50\%) escape fractions could contribute to the low EW as well. 
Conversely, we stress again that a population of young (age $\leq$ 10 Myr), blue ($\beta \sim -2$) galaxies with higher sSFR exists among all density regimes, indicating that star formation is actively proceeding in all areas of the protocluster, as it should according to simulations \citep{Chiang2013a, Rennehan2024a}.

Lastly, none of the galaxies presented here is detected in dust continuum in the 1.1$\times$1.1 arcmin$^2$ ALMA Band 6 mosaic around J0305 (Wang et al., in prep.). 
Therefore, we can conclude that the protocluster is not undergoing a period of simultaneous dusty starbursts as is common in protoclusters at lower ($2<z<4$) redshift \citep[e.g.,][]{Dannerbauer2014a, Casey2015a, Harikane2019a, Hill2020a, Long2020a}. 
Alternatively, more massive galaxies with $M_*>$10$^9$\,\msun\, could be moderately dust obscured enough not to show up in this sample but still too faint in the IR to be picked up by ALMA, so we would need deeper photometry with NIRCam/MIRI \citep[e.g.,][]{Shivaei2024a}.
Regardless, if those protoclusters at $2<z<4$ are descendant structures from overdensities such as the ones presented here, the UV-bright star formation in a rich population of primarily low-mass galaxies at $z>6$ marks an important evolutionary stage predating the massive, dusty star-forming phase characterized by frequent mergers.

\section{Conclusions}\label{sec:conclusion}

In this work we have characterized a sample of 124 \Oiii\, emitters identified in the ASPIRE grism+imaging quasar legacy survey over an area of 35 arcmin$^2$.
53 of these galaxies are members of a quasar-anchored protocluster at $z=6.6$, while 18 and 20 galaxies occupy serendipitously discovered overdensities at $z=6.2$ and $z=5.4$. 
The remaining galaxies served as a field sample as a comparison between the evolution of galaxies within and without overdensities during the epoch of reionization.
We performed SED fitting and spectral analysis to derive galaxy properties such as stellar mass, age, and star formation rates, as well as search for signatures of AGN activity.

In a spatial analysis of the protocluster, we found that:

\begin{itemize}
\item Using \texttt{Prospector} as our fiducial model to fit the SEDs of the protocluster sample, we compute a median stellar mass of log($M_*/M_{\odot}$) = 8.31 $\pm$ 0.08, median SFR 4.6 $\pm$ 0.7 M$_{\odot}$\,yr$^{-1}$, and median age of 23 $\pm$ 1 Myr.
These are, respectively, 0.30, 0.20, and 0.35 dex higher than in the field sample, significant at a confidence level $p<0.05$.

\item We searched for AGN using broad-line criteria, the mass-excitation diagram, and ``little red dot" color/compactness criteria. There are no galaxies with evidence of broad H$\beta$, with a stacked average FWHM of 585 $\pm$ 152\,km\,s$^{-1}$. 
Five galaxies \textit{both} lie above the star formation boundary of the MEx diagram and meet the color/compactness criteria of a little red dot, so we tentatively assume these are Type 2 AGN candidates.
Four of these galaxies are within the protocluster and they are all located on the outskirts of the filaments, implying a physical mechanism that could spark AGN acitivity due to recent gas inflow.

\item There is no correlation between any of our derived SED and separation from the quasar, i.e. the quasar host galaxy is not predominantly affecting galaxy evolution in its immediate vicinity. 
However, we found a mild positive correlation between \Oiii+H$\beta$ EW and distance from the quasar, implying that recent star formation has ramped up in the filaments further away from the quasar.

\item We found that there is a mild positive correlation between local spatial density $\Sigma_3$ and stellar mass as well as stellar age, and a significantly negative correlation with the burstiness parameter SFR$_{10 \rm Myr}$/SFR$_{100\rm Myr}$ and \Oiii+H$\beta$ EW. This implies that the massive galaxies in denser filaments formed their stars earlier, whereas star formation is ramping up in lower-mass galaxies on the outskirts of the protocluster.

\item In a 3D representation of the protocluster, we see evidence that the most massive, oldest galaxies are located along dense filaments which are surrounded on their outskirts by younger galaxies with higher specific star formation rates. We conclude that the dense environment does affect galaxy evolution in an inside-out fashion.

\end{itemize}

Overall, we find significant evidence for environmental effects in early protocluster filaments, predating more extreme bursts of star formation in descendant structures.
A larger quasar sample will be presented in future ASPIRE publications which will help build up population statistics for quasar environments.


\begin{acknowledgments}

The authors thank Jakob Helton, Lily Whitler and Steven Finkelstein for helpful discussions and advice on SED fitting.
The authors also thank Kazuaki Ota for kindly providing the Subaru data used in this work.
JBC  acknowledges funding from the JWST Arizona/Steward Postdoc in Early galaxies and Reionization (JASPER) Scholar contract at the University of Arizona. 
FW and JBC acknowledge support from NSF Grant AST-2308258.
CM acknowledges support from Fondecyt Iniciacion grant 11240336  and ANID BASAL project FB210003.
FL acknowledges support from the INAF GO 2022 grant ``The birth of the giants: JWST sheds light on the build-up of quasars at cosmic dawn" and from the INAF 2023 mini-grant ``Exploiting the powerful capabilities of JWST/NIRSpec to unveil the distant Universe."
AL acknowledges support by the PRIN MUR ``2022935STW."
SZ acknowledges support from the National Science Foundation of China (grant no.\ 12303011).

This work is based on observations made with the NASA/ESA/CSA James Webb Space Telescope. The data were obtained from the Mikulski Archive for Space Telescopes at the Space Telescope Science Institute, which is operated by the Association of Universities for Research in Astronomy, Inc., under NASA contract NAS 5-03127 for JWST. These observations are associated with programs \#2078 and \#3225. Support for these programs was given through a grant from the Space Telescope Science Institute, which is operated by the Association of Universities for Research in Astronomy, Inc., under NASA contract NAS 5-03127.

\software{Astropy \citep{astropy:2013, astropy:2018, astropy:2022}, SourceXtractor++ \citep{Bertin2020a}}

\end{acknowledgments}

\bibliography{main}

\begin{thebibliography}{}
\expandafter\ifx\csname natexlab\endcsname\relax\def\natexlab#1{#1}\fi
\providecommand{\url}[1]{\href{#1}{#1}}
\providecommand{\dodoi}[1]{doi:~\href{http://doi.org/#1}{\nolinkurl{#1}}}
\providecommand{\doeprint}[1]{\href{http://ascl.net/#1}{\nolinkurl{http://ascl.net/#1}}}
\providecommand{\doarXiv}[1]{\href{https://arxiv.org/abs/#1}{\nolinkurl{https://arxiv.org/abs/#1}}}

\bibitem[{{Akins} {et~al.}(2024){Akins}, {Casey}, {Lambrides}, {Allen},
  {Andika}, {Brinch}, {Champagne}, {Cooper}, {Ding}, {Drakos}, {Faisst},
  {Finkelstein}, {Franco}, {Fujimoto}, {Gentile}, {Gillman}, {Gozaliasl},
  {Harish}, {Hayward}, {Hirschmann}, {Ilbert}, {Kartaltepe}, {Kocevski},
  {Koekemoer}, {Kokorev}, {Liu}, {Long}, {McCracken}, {McKinney}, {Onoue},
  {Paquereau}, {Renzini}, {Rhodes}, {Robertson}, {Shuntov}, {Silverman},
  {Tanaka}, {Toft}, {Trakhtenbrot}, {Valentino}, \& {Zavala}}]{Akins2024a}
{Akins}, H.~B., {Casey}, C.~M., {Lambrides}, E., {et~al.} 2024, arXiv e-prints,
  arXiv:2406.10341, \dodoi{10.48550/arXiv.2406.10341}

\bibitem[{{Arellano-C{\'o}rdova} {et~al.}(2022){Arellano-C{\'o}rdova}, {Berg},
  {Chisholm}, {Arrabal Haro}, {Dickinson}, {Finkelstein}, {Leclercq}, {Rogers},
  {Simons}, {Skillman}, {Trump}, \& {Kartaltepe}}]{ArellanoCordova2022a}
{Arellano-C{\'o}rdova}, K.~Z., {Berg}, D.~A., {Chisholm}, J., {et~al.} 2022,
  \apjl, 940, L23, \dodoi{10.3847/2041-8213/ac9ab2}

\bibitem[{{Astropy Collaboration} {et~al.}(2013){Astropy Collaboration},
  {Robitaille}, {Tollerud}, {Greenfield}, {Droettboom}, {Bray}, {Aldcroft},
  {Davis}, {Ginsburg}, {Price-Whelan}, {Kerzendorf}, {Conley}, {Crighton},
  {Barbary}, {Muna}, {Ferguson}, {Grollier}, {Parikh}, {Nair}, {Unther},
  {Deil}, {Woillez}, {Conseil}, {Kramer}, {Turner}, {Singer}, {Fox}, {Weaver},
  {Zabalza}, {Edwards}, {Azalee Bostroem}, {Burke}, {Casey}, {Crawford},
  {Dencheva}, {Ely}, {Jenness}, {Labrie}, {Lim}, {Pierfederici}, {Pontzen},
  {Ptak}, {Refsdal}, {Servillat}, \& {Streicher}}]{astropy:2013}
{Astropy Collaboration}, {Robitaille}, T.~P., {Tollerud}, E.~J., {et~al.} 2013,
  \aap, 558, A33, \dodoi{10.1051/0004-6361/201322068}

\bibitem[{{Astropy Collaboration} {et~al.}(2018){Astropy Collaboration},
  {Price-Whelan}, {Sip{\H{o}}cz}, {G{\"u}nther}, {Lim}, {Crawford}, {Conseil},
  {Shupe}, {Craig}, {Dencheva}, {Ginsburg}, {Vand erPlas}, {Bradley},
  {P{\'e}rez-Su{\'a}rez}, {de Val-Borro}, {Aldcroft}, {Cruz}, {Robitaille},
  {Tollerud}, {Ardelean}, {Babej}, {Bach}, {Bachetti}, {Bakanov}, {Bamford},
  {Barentsen}, {Barmby}, {Baumbach}, {Berry}, {Biscani}, {Boquien}, {Bostroem},
  {Bouma}, {Brammer}, {Bray}, {Breytenbach}, {Buddelmeijer}, {Burke},
  {Calderone}, {Cano Rodr{\'\i}guez}, {Cara}, {Cardoso}, {Cheedella}, {Copin},
  {Corrales}, {Crichton}, {D'Avella}, {Deil}, {Depagne}, {Dietrich}, {Donath},
  {Droettboom}, {Earl}, {Erben}, {Fabbro}, {Ferreira}, {Finethy}, {Fox},
  {Garrison}, {Gibbons}, {Goldstein}, {Gommers}, {Greco}, {Greenfield},
  {Groener}, {Grollier}, {Hagen}, {Hirst}, {Homeier}, {Horton}, {Hosseinzadeh},
  {Hu}, {Hunkeler}, {Ivezi{\'c}}, {Jain}, {Jenness}, {Kanarek}, {Kendrew},
  {Kern}, {Kerzendorf}, {Khvalko}, {King}, {Kirkby}, {Kulkarni}, {Kumar},
  {Lee}, {Lenz}, {Littlefair}, {Ma}, {Macleod}, {Mastropietro}, {McCully},
  {Montagnac}, {Morris}, {Mueller}, {Mumford}, {Muna}, {Murphy}, {Nelson},
  {Nguyen}, {Ninan}, {N{\"o}the}, {Ogaz}, {Oh}, {Parejko}, {Parley}, {Pascual},
  {Patil}, {Patil}, {Plunkett}, {Prochaska}, {Rastogi}, {Reddy Janga},
  {Sabater}, {Sakurikar}, {Seifert}, {Sherbert}, {Sherwood-Taylor}, {Shih},
  {Sick}, {Silbiger}, {Singanamalla}, {Singer}, {Sladen}, {Sooley},
  {Sornarajah}, {Streicher}, {Teuben}, {Thomas}, {Tremblay}, {Turner},
  {Terr{\'o}n}, {van Kerkwijk}, {de la Vega}, {Watkins}, {Weaver}, {Whitmore},
  {Woillez}, {Zabalza}, \& {Astropy Contributors}}]{astropy:2018}
{Astropy Collaboration}, {Price-Whelan}, A.~M., {Sip{\H{o}}cz}, B.~M., {et~al.}
  2018, \aj, 156, 123, \dodoi{10.3847/1538-3881/aabc4f}

\bibitem[{{Astropy Collaboration} {et~al.}(2022){Astropy Collaboration},
  {Price-Whelan}, {Lim}, {Earl}, {Starkman}, {Bradley}, {Shupe}, {Patil},
  {Corrales}, {Brasseur}, {N{"o}the}, {Donath}, {Tollerud}, {Morris},
  {Ginsburg}, {Vaher}, {Weaver}, {Tocknell}, {Jamieson}, {van Kerkwijk},
  {Robitaille}, {Merry}, {Bachetti}, {G{"u}nther}, {Aldcroft},
  {Alvarado-Montes}, {Archibald}, {B{'o}di}, {Bapat}, {Barentsen}, {Baz{'a}n},
  {Biswas}, {Boquien}, {Burke}, {Cara}, {Cara}, {Conroy}, {Conseil}, {Craig},
  {Cross}, {Cruz}, {D'Eugenio}, {Dencheva}, {Devillepoix}, {Dietrich},
  {Eigenbrot}, {Erben}, {Ferreira}, {Foreman-Mackey}, {Fox}, {Freij}, {Garg},
  {Geda}, {Glattly}, {Gondhalekar}, {Gordon}, {Grant}, {Greenfield}, {Groener},
  {Guest}, {Gurovich}, {Handberg}, {Hart}, {Hatfield-Dodds}, {Homeier},
  {Hosseinzadeh}, {Jenness}, {Jones}, {Joseph}, {Kalmbach}, {Karamehmetoglu},
  {Ka{l}uszy{'n}ski}, {Kelley}, {Kern}, {Kerzendorf}, {Koch}, {Kulumani},
  {Lee}, {Ly}, {Ma}, {MacBride}, {Maljaars}, {Muna}, {Murphy}, {Norman},
  {O'Steen}, {Oman}, {Pacifici}, {Pascual}, {Pascual-Granado}, {Patil},
  {Perren}, {Pickering}, {Rastogi}, {Roulston}, {Ryan}, {Rykoff}, {Sabater},
  {Sakurikar}, {Salgado}, {Sanghi}, {Saunders}, {Savchenko}, {Schwardt},
  {Seifert-Eckert}, {Shih}, {Jain}, {Shukla}, {Sick}, {Simpson},
  {Singanamalla}, {Singer}, {Singhal}, {Sinha}, {Sip{H{o}}cz}, {Spitler},
  {Stansby}, {Streicher}, {{{S}}umak}, {Swinbank}, {Taranu}, {Tewary},
  {Tremblay}, {Val-Borro}, {Van Kooten}, {Vasovi{'c}}, {Verma}, {de Miranda
  Cardoso}, {Williams}, {Wilson}, {Winkel}, {Wood-Vasey}, {Xue}, {Yoachim},
  {Zhang}, {Zonca}, \& {Astropy Project Contributors}}]{astropy:2022}
{Astropy Collaboration}, {Price-Whelan}, A.~M., {Lim}, P.~L., {et~al.} 2022,
  \apj, 935, 167, \dodoi{10.3847/1538-4357/ac7c74}

\bibitem[{{Ba{\~n}ados} {et~al.}(2013){Ba{\~n}ados}, {Venemans}, {Walter},
  {Kurk}, {Overzier}, \& {Ouchi}}]{Banados2013a}
{Ba{\~n}ados}, E., {Venemans}, B., {Walter}, F., {et~al.} 2013, \apj, 773, 178,
  \dodoi{10.1088/0004-637X/773/2/178}

\bibitem[{{Bassini} {et~al.}(2020){Bassini}, {Rasia}, {Borgani}, {Granato},
  {Ragone-Figueroa}, {Biffi}, {Ragagnin}, {Dolag}, {Lin}, {Murante},
  {Napolitano}, {Taffoni}, {Tornatore}, \& {Wang}}]{Bassini2020a}
{Bassini}, L., {Rasia}, E., {Borgani}, S., {et~al.} 2020, \aap, 642, A37,
  \dodoi{10.1051/0004-6361/202038396}

\bibitem[{{Bertin} {et~al.}(2020){Bertin}, {Schefer}, {Apostolakos},
  {{\'A}lvarez-Ayll{\'o}n}, {Dubath}, \& {K{\"u}mmel}}]{Bertin2020a}
{Bertin}, E., {Schefer}, M., {Apostolakos}, N., {et~al.} 2020, in Astronomical
  Society of the Pacific Conference Series, Vol. 527, Astronomical Data
  Analysis Software and Systems XXIX, ed. R.~{Pizzo}, E.~R. {Deul}, J.~D.
  {Mol}, J.~{de Plaa}, \& H.~{Verkouter}, 461

\bibitem[{{Bian} {et~al.}(2018){Bian}, {Kewley}, \& {Dopita}}]{Bian2018a}
{Bian}, F., {Kewley}, L.~J., \& {Dopita}, M.~A. 2018, \apj, 859, 175,
  \dodoi{10.3847/1538-4357/aabd74}

\bibitem[{{Boyett} {et~al.}(2024){Boyett}, {Bunker}, {Curtis-Lake},
  {Chevallard}, {Cameron}, {Jones}, {Saxena}, {Charlot}, {Curti}, {Wallace},
  {Arribas}, {Carniani}, {Willott}, {Alberts}, {Eisenstein}, {Hainline},
  {Hausen}, {Johnson}, {Rieke}, {Robertson}, {Stark}, {Tacchella}, {Williams},
  {Chen}, {Egami}, {Endsley}, {Kumari}, {Laseter}, {Looser}, {Maseda},
  {Scholtz}, {Shivaei}, {Simmonds}, {Smit}, {{\"U}bler}, \&
  {Witstok}}]{Boyett2024a}
{Boyett}, K., {Bunker}, A.~J., {Curtis-Lake}, E., {et~al.} 2024, \mnras, 535,
  1796, \dodoi{10.1093/mnras/stae2430}

\bibitem[{{Brammer} {et~al.}(2008){Brammer}, {van Dokkum}, \&
  {Coppi}}]{Brammer2008a}
{Brammer}, G.~B., {van Dokkum}, P.~G., \& {Coppi}, P. 2008, \apj, 686, 1503,
  \dodoi{10.1086/591786}

\bibitem[{{Bruzual} \& {Charlot}(2003)}]{Bruzual2003a}
{Bruzual}, G., \& {Charlot}, S. 2003, \mnras, 344, 1000,
  \dodoi{10.1046/j.1365-8711.2003.06897.x}

\bibitem[{{Carnall} {et~al.}(2019){Carnall}, {Leja}, {Johnson}, {McLure},
  {Dunlop}, \& {Conroy}}]{Carnall2019a}
{Carnall}, A.~C., {Leja}, J., {Johnson}, B.~D., {et~al.} 2019, \apj, 873, 44,
  \dodoi{10.3847/1538-4357/ab04a2}

\bibitem[{{Carnall} {et~al.}(2018){Carnall}, {McLure}, {Dunlop}, \&
  {Dav{\'e}}}]{Carnall2018a}
{Carnall}, A.~C., {McLure}, R.~J., {Dunlop}, J.~S., \& {Dav{\'e}}, R. 2018,
  \mnras, 480, 4379, \dodoi{10.1093/mnras/sty2169}

\bibitem[{{Casey} {et~al.}(2015){Casey}, {Cooray}, {Capak}, {Fu}, {Kovac},
  {Lilly}, {Sanders}, {Scoville}, \& {Treister}}]{Casey2015a}
{Casey}, C.~M., {Cooray}, A., {Capak}, P., {et~al.} 2015, \apjl, 808, L33,
  \dodoi{10.1088/2041-8205/808/2/L33}

\bibitem[{{Champagne} {et~al.}(2023){Champagne}, {Casey}, {Finkelstein},
  {Bagley}, {Cooper}, {Larson}, {Long}, \& {Wang}}]{Champagne2023a}
{Champagne}, J.~B., {Casey}, C.~M., {Finkelstein}, S.~L., {et~al.} 2023, \apj,
  952, 99, \dodoi{10.3847/1538-4357/acda8d}

\bibitem[{{Champagne} {et~al.}(2018){Champagne}, {Decarli}, {Casey},
  {Venemans}, {Ba{\~n}ados}, {Walter}, {Bertoldi}, {Fan}, {Farina},
  {Mazzucchelli}, {Riechers}, {Strauss}, {Wang}, \& {Yang}}]{Champagne2018a}
{Champagne}, J.~B., {Decarli}, R., {Casey}, C.~M., {et~al.} 2018, \apj, 867,
  153, \dodoi{10.3847/1538-4357/aae396}

\bibitem[{{Champagne} {et~al.}(2021){Champagne}, {Casey}, {Zavala}, {Cooray},
  {Dannerbauer}, {Fabian}, {Hayward}, {Long}, \& {Spilker}}]{Champagne2021a}
{Champagne}, J.~B., {Casey}, C.~M., {Zavala}, J.~A., {et~al.} 2021, \apj, 913,
  110, \dodoi{10.3847/1538-4357/abf4e6}

\bibitem[{{Champagne} {et~al.}(2024){Champagne}, {Wang}, {Zhang}, {Yang},
  {Fan}, {Hennawi}, {Sun}, {Ba{\~n}ados}, {Bosman}, {Costa}, {Eilers},
  {Endsley}, {Jin}, {Jun}, {Li}, {Lin}, {Liu}, {Loiacono}, {Lupi},
  {Mazzucchelli}, {Pudoka}, {Protu{\v{s}}ov{\`a}}, {Rojas-Ruiz}, {Tee},
  {Trebitsch}, {Venemans}, {Zhuang}, \& {Zou}}]{Champagne2024a}
{Champagne}, J.~B., {Wang}, F., {Zhang}, H., {et~al.} 2024, arXiv e-prints,
  arXiv:2410.03826, \dodoi{10.48550/arXiv.2410.03826}

\bibitem[{{Chiang} {et~al.}(2013){Chiang}, {Overzier}, \&
  {Gebhardt}}]{Chiang2013a}
{Chiang}, Y.-K., {Overzier}, R., \& {Gebhardt}, K. 2013, \apj, 779, 127,
  \dodoi{10.1088/0004-637X/779/2/127}

\bibitem[{{Chiang} {et~al.}(2017){Chiang}, {Overzier}, {Gebhardt}, \&
  {Henriques}}]{Chiang2017a}
{Chiang}, Y.-K., {Overzier}, R.~A., {Gebhardt}, K., \& {Henriques}, B. 2017,
  \apjl, 844, L23, \dodoi{10.3847/2041-8213/aa7e7b}

\bibitem[{{Choi} {et~al.}(2016){Choi}, {Dotter}, {Conroy}, {Cantiello},
  {Paxton}, \& {Johnson}}]{Choi2016a}
{Choi}, J., {Dotter}, A., {Conroy}, C., {et~al.} 2016, \apj, 823, 102,
  \dodoi{10.3847/0004-637X/823/2/102}

\bibitem[{{Conroy} {et~al.}(2009){Conroy}, {Gunn}, \& {White}}]{Conroy2009a}
{Conroy}, C., {Gunn}, J.~E., \& {White}, M. 2009, \apj, 699, 486,
  \dodoi{10.1088/0004-637X/699/1/486}

\bibitem[{{Costa} {et~al.}(2014){Costa}, {Sijacki}, {Trenti}, \&
  {Haehnelt}}]{Costa2014a}
{Costa}, T., {Sijacki}, D., {Trenti}, M., \& {Haehnelt}, M.~G. 2014, \mnras,
  439, 2146, \dodoi{10.1093/mnras/stu101}

\bibitem[{{Dannerbauer} {et~al.}(2014){Dannerbauer}, {Kurk}, {De Breuck},
  {Wylezalek}, {Santos}, {Koyama}, {Seymour}, {Tanaka}, {Hatch}, {Altieri},
  {Coia}, {Galametz}, {Kodama}, {Miley}, {R{\"o}ttgering}, {Sanchez-Portal},
  {Valtchanov}, {Venemans}, \& {Ziegler}}]{Dannerbauer2014a}
{Dannerbauer}, H., {Kurk}, J.~D., {De Breuck}, C., {et~al.} 2014, \aap, 570,
  A55, \dodoi{10.1051/0004-6361/201423771}

\bibitem[{{Dotter}(2016)}]{Dotter2016a}
{Dotter}, A. 2016, \apjs, 222, 8, \dodoi{10.3847/0067-0049/222/1/8}

\bibitem[{{Eilers} {et~al.}(2017){Eilers}, {Davies}, {Hennawi}, {Prochaska},
  {Luki{\'c}}, \& {Mazzucchelli}}]{Eilers2017a}
{Eilers}, A.-C., {Davies}, F.~B., {Hennawi}, J.~F., {et~al.} 2017, \apj, 840,
  24, \dodoi{10.3847/1538-4357/aa6c60}

\bibitem[{{Eilers} {et~al.}(2024){Eilers}, {Mackenzie}, {Pizzati}, {Matthee},
  {Hennawi}, {Zhang}, {Bordoloi}, {Kashino}, {Lilly}, {Naidu}, {Simcoe}, {Yue},
  {Frenk}, {Helly}, {Schaller}, \& {Schaye}}]{Eilers2024a}
{Eilers}, A.-C., {Mackenzie}, R., {Pizzati}, E., {et~al.} 2024, \apj, 974, 275,
  \dodoi{10.3847/1538-4357/ad778b}

\bibitem[{{Eldridge} \& {Stanway}(2022)}]{Eldridge2022a}
{Eldridge}, J.~J., \& {Stanway}, E.~R. 2022, \araa, 60, 455,
  \dodoi{10.1146/annurev-astro-052920-100646}

\bibitem[{{Endsley} {et~al.}(2021){Endsley}, {Stark}, {Chevallard}, \&
  {Charlot}}]{Endsley2021a}
{Endsley}, R., {Stark}, D.~P., {Chevallard}, J., \& {Charlot}, S. 2021, \mnras,
  500, 5229, \dodoi{10.1093/mnras/staa3370}

\bibitem[{{Endsley} {et~al.}(2023{\natexlab{a}}){Endsley}, {Stark}, {Whitler},
  {Topping}, {Chen}, {Plat}, {Chisholm}, \& {Charlot}}]{Endsley2023a}
{Endsley}, R., {Stark}, D.~P., {Whitler}, L., {et~al.} 2023{\natexlab{a}},
  \mnras, 524, 2312, \dodoi{10.1093/mnras/stad1919}

\bibitem[{{Endsley} {et~al.}(2023{\natexlab{b}}){Endsley}, {Stark}, {Whitler},
  {Topping}, {Johnson}, {Robertson}, {Tacchella}, {Alberts}, {Baker},
  {Bhatawdekar}, {Boyett}, {Bunker}, {Cameron}, {Carniani}, {Charlot}, {Chen},
  {Chevallard}, {Curtis-Lake}, {Danhaive}, {Egami}, {Eisenstein}, {Hainline},
  {Helton}, {Ji}, {Looser}, {Maiolino}, {Nelson}, {Pusk{\'a}s}, {Rieke},
  {Rieke}, {Rix}, {Sandles}, {Saxena}, {Simmonds}, {Smit}, {Sun}, {Williams},
  {Willmer}, {Willott}, \& {Witstok}}]{Endsley2023b}
---. 2023{\natexlab{b}}, arXiv e-prints, arXiv:2306.05295,
  \dodoi{10.48550/arXiv.2306.05295}

\bibitem[{{Falc{\'o}n-Barroso} {et~al.}(2011){Falc{\'o}n-Barroso},
  {S{\'a}nchez-Bl{\'a}zquez}, {Vazdekis}, {Ricciardelli}, {Cardiel}, {Cenarro},
  {Gorgas}, \& {Peletier}}]{Falcon-Barroso2011a}
{Falc{\'o}n-Barroso}, J., {S{\'a}nchez-Bl{\'a}zquez}, P., {Vazdekis}, A.,
  {et~al.} 2011, \aap, 532, A95, \dodoi{10.1051/0004-6361/201116842}

\bibitem[{{Fan} {et~al.}(2023){Fan}, {Ba{\~n}ados}, \& {Simcoe}}]{Fan2023a}
{Fan}, X., {Ba{\~n}ados}, E., \& {Simcoe}, R.~A. 2023, \araa, 61, 373,
  \dodoi{10.1146/annurev-astro-052920-102455}

\bibitem[{{Farina} {et~al.}(2017){Farina}, {Venemans}, {Decarli}, {Hennawi},
  {Walter}, {Ba{\~n}ados}, {Mazzucchelli}, {Cantalupo}, {Arrigoni-Battaia}, \&
  {McGreer}}]{Farina2017a}
{Farina}, E.~P., {Venemans}, B.~P., {Decarli}, R., {et~al.} 2017, \apj, 848,
  78, \dodoi{10.3847/1538-4357/aa8df4}

\bibitem[{{Ferland} {et~al.}(2017){Ferland}, {Chatzikos}, {Guzm{\'a}n},
  {Lykins}, {van Hoof}, {Williams}, {Abel}, {Badnell}, {Keenan}, {Porter}, \&
  {Stancil}}]{Ferland2017a}
{Ferland}, G.~J., {Chatzikos}, M., {Guzm{\'a}n}, F., {et~al.} 2017, \rmxaa, 53,
  385, \dodoi{10.48550/arXiv.1705.10877}

\bibitem[{{Feroz} {et~al.}(2019){Feroz}, {Hobson}, {Cameron}, \&
  {Pettitt}}]{Feroz2019a}
{Feroz}, F., {Hobson}, M.~P., {Cameron}, E., \& {Pettitt}, A.~N. 2019, The Open
  Journal of Astrophysics, 2, 10, \dodoi{10.21105/astro.1306.2144}

\bibitem[{{Finkelstein} {et~al.}(2015){Finkelstein}, {Ryan}, {Papovich},
  {Dickinson}, {Song}, {Somerville}, {Ferguson}, {Salmon}, {Giavalisco},
  {Koekemoer}, {Ashby}, {Behroozi}, {Castellano}, {Dunlop}, {Faber}, {Fazio},
  {Fontana}, {Grogin}, {Hathi}, {Jaacks}, {Kocevski}, {Livermore}, {McLure},
  {Merlin}, {Mobasher}, {Newman}, {Rafelski}, {Tilvi}, \&
  {Willner}}]{Finkelstein2015a}
{Finkelstein}, S.~L., {Ryan}, Russell~E., J., {Papovich}, C., {et~al.} 2015,
  \apj, 810, 71, \dodoi{10.1088/0004-637X/810/1/71}

\bibitem[{{Forrest} {et~al.}(2024){Forrest}, {Lemaux}, {Shah}, {Staab}, {Gal},
  {Lubin}, {Cooper}, {Cucciati}, {Hung}, {McConachie}, {Muzzin}, {Wilson},
  {Bardelli}, {Cassar{\`a}}, {Chang}, {Giddings}, {Golden-Marx}, {Hathi},
  {Urbano Stawinski}, \& {Zucca}}]{Forrest2024a}
{Forrest}, B., {Lemaux}, B.~C., {Shah}, E.~A., {et~al.} 2024, \apj, 971, 169,
  \dodoi{10.3847/1538-4357/ad5e78}

\bibitem[{{Greene} {et~al.}(2024){Greene}, {Labbe}, {Goulding}, {Furtak},
  {Chemerynska}, {Kokorev}, {Dayal}, {Volonteri}, {Williams}, {Wang}, {Setton},
  {Burgasser}, {Bezanson}, {Atek}, {Brammer}, {Cutler}, {Feldmann}, {Fujimoto},
  {Glazebrook}, {de Graaff}, {Khullar}, {Leja}, {Marchesini}, {Maseda},
  {Matthee}, {Miller}, {Naidu}, {Nanayakkara}, {Oesch}, {Pan}, {Papovich},
  {Price}, {van Dokkum}, {Weaver}, {Whitaker}, \& {Zitrin}}]{Greene2024a}
{Greene}, J.~E., {Labbe}, I., {Goulding}, A.~D., {et~al.} 2024, \apj, 964, 39,
  \dodoi{10.3847/1538-4357/ad1e5f}

\bibitem[{{Harikane} {et~al.}(2019){Harikane}, {Ouchi}, {Ono}, {Fujimoto},
  {Donevski}, {Shibuya}, {Faisst}, {Goto}, {Hatsukade}, {Kashikawa}, {Kohno},
  {Hashimoto}, {Higuchi}, {Inoue}, {Lin}, {Martin}, {Overzier}, {Smail},
  {Toshikawa}, {Umehata}, {Ao}, {Chapman}, {Clements}, {Im}, {Jing},
  {Kawaguchi}, {Lee}, {Lee}, {Lin}, {Matsuoka}, {Marinello}, {Nagao},
  {Onodera}, {Toft}, \& {Wang}}]{Harikane2019a}
{Harikane}, Y., {Ouchi}, M., {Ono}, Y., {et~al.} 2019, \apj, 883, 142,
  \dodoi{10.3847/1538-4357/ab2cd5}

\bibitem[{{Harikane} {et~al.}(2023){Harikane}, {Ouchi}, {Oguri}, {Ono},
  {Nakajima}, {Isobe}, {Umeda}, {Mawatari}, \& {Zhang}}]{Harikane2023a}
{Harikane}, Y., {Ouchi}, M., {Oguri}, M., {et~al.} 2023, \apjs, 265, 5,
  \dodoi{10.3847/1538-4365/acaaa9}

\bibitem[{{Helton} {et~al.}(2024){Helton}, {Sun}, {Woodrum}, {Hainline},
  {Willmer}, {Rieke}, {Rieke}, {Tacchella}, {Robertson}, {Johnson}, {Alberts},
  {Eisenstein}, {Hausen}, {Bonaventura}, {Bunker}, {Charlot}, {Curti},
  {Curtis-Lake}, {Looser}, {Maiolino}, {Willott}, {Witstok}, {Boyett}, {Chen},
  {Egami}, {Endsley}, {Hviding}, {Jaffe}, {Ji}, {Lyu}, \&
  {Sandles}}]{Helton2024a}
{Helton}, J.~M., {Sun}, F., {Woodrum}, C., {et~al.} 2024, \apj, 962, 124,
  \dodoi{10.3847/1538-4357/ad0da7}

\bibitem[{{Herard-Demanche} {et~al.}(2025){Herard-Demanche}, {Bouwens},
  {Oesch}, {Naidu}, {Decarli}, {Nelson}, {Brammer}, {Weibel}, {Xiao},
  {Stefanon}, {Walter}, {Matthee}, {Meyer}, {Wuyts}, {Reddy}, {Rowland}, {van
  Leeuwen}, {Haro}, {Dannerbauer}, {Shapley}, {Chisholm}, {van Dokkum},
  {Labbe}, {Illingworth}, {Schaerer}, \& {Shivaei}}]{Herard-Demanche2025a}
{Herard-Demanche}, T., {Bouwens}, R.~J., {Oesch}, P.~A., {et~al.} 2025, \mnras,
  537, 788, \dodoi{10.1093/mnras/staf030}

\bibitem[{{Hill} {et~al.}(2020){Hill}, {Chapman}, {Scott}, {Apostolovski},
  {Aravena}, {B{\'e}thermin}, {Bradford}, {Canning}, {De Breuck}, {Dong},
  {Gonzalez}, {Greve}, {Hayward}, {Hezaveh}, {Litke}, {Malkan}, {Marrone},
  {Phadke}, {Reuter}, {Rotermund}, {Spilker}, {Vieira}, \&
  {Wei{\ss}}}]{Hill2020a}
{Hill}, R., {Chapman}, S., {Scott}, D., {et~al.} 2020, \mnras, 495, 3124,
  \dodoi{10.1093/mnras/staa1275}

\bibitem[{{Hirschmann} {et~al.}(2023){Hirschmann}, {Charlot}, {Feltre},
  {Curtis-Lake}, {Somerville}, {Chevallard}, {Choi}, {Nelson}, {Morisset},
  {Plat}, \& {Vidal-Garcia}}]{Hirschmann2023a}
{Hirschmann}, M., {Charlot}, S., {Feltre}, A., {et~al.} 2023, \mnras, 526,
  3610, \dodoi{10.1093/mnras/stad2955}

\bibitem[{{Inoue} {et~al.}(2014){Inoue}, {Shimizu}, {Iwata}, \&
  {Tanaka}}]{Inoue2014a}
{Inoue}, A.~K., {Shimizu}, I., {Iwata}, I., \& {Tanaka}, M. 2014, \mnras, 442,
  1805, \dodoi{10.1093/mnras/stu936}

\bibitem[{{Johnson} {et~al.}(2019){Johnson}, {Leja}, {Conroy}, \&
  {Speagle}}]{Johnson2019a}
{Johnson}, B.~D., {Leja}, J.~L., {Conroy}, C., \& {Speagle}, J.~S. 2019,
  {Prospector: Stellar population inference from spectra and SEDs},
  Astrophysics Source Code Library, record ascl:1905.025.
\newblock \doeprint{1905.025}

\bibitem[{{Juneau} {et~al.}(2011){Juneau}, {Dickinson}, {Alexander}, \&
  {Salim}}]{Juneau2011a}
{Juneau}, S., {Dickinson}, M., {Alexander}, D.~M., \& {Salim}, S. 2011, \apj,
  736, 104, \dodoi{10.1088/0004-637X/736/2/104}

\bibitem[{{Juneau} {et~al.}(2014){Juneau}, {Bournaud}, {Charlot}, {Daddi},
  {Elbaz}, {Trump}, {Brinchmann}, {Dickinson}, {Duc}, {Gobat}, {Jean-Baptiste},
  {Le Floc'h}, {Lehnert}, {Pacifici}, {Pannella}, \& {Schreiber}}]{Juneau2014a}
{Juneau}, S., {Bournaud}, F., {Charlot}, S., {et~al.} 2014, \apj, 788, 88,
  \dodoi{10.1088/0004-637X/788/1/88}

\bibitem[{{Kashino} {et~al.}(2023){Kashino}, {Lilly}, {Matthee}, {Eilers},
  {Mackenzie}, {Bordoloi}, \& {Simcoe}}]{Kashino2023a}
{Kashino}, D., {Lilly}, S.~J., {Matthee}, J., {et~al.} 2023, \apj, 950, 66,
  \dodoi{10.3847/1538-4357/acc588}

\bibitem[{{Kim} {et~al.}(2009){Kim}, {Stiavelli}, {Trenti}, {Pavlovsky},
  {Djorgovski}, {Scarlata}, {Stern}, {Mahabal}, {Thompson}, {Dickinson},
  {Panagia}, \& {Meylan}}]{Kim2009a}
{Kim}, S., {Stiavelli}, M., {Trenti}, M., {et~al.} 2009, \apj, 695, 809,
  \dodoi{10.1088/0004-637X/695/2/809}

\bibitem[{{Kokorev} {et~al.}(2024){Kokorev}, {Caputi}, {Greene}, {Dayal},
  {Trebitsch}, {Cutler}, {Fujimoto}, {Labb{\'e}}, {Miller}, {Iani},
  {Navarro-Carrera}, \& {Rinaldi}}]{Kokorev2024a}
{Kokorev}, V., {Caputi}, K.~I., {Greene}, J.~E., {et~al.} 2024, \apj, 968, 38,
  \dodoi{10.3847/1538-4357/ad4265}

\bibitem[{{Kravtsov} \& {Borgani}(2012)}]{Kravtsov2012a}
{Kravtsov}, A.~V., \& {Borgani}, S. 2012, \araa, 50, 353,
  \dodoi{10.1146/annurev-astro-081811-125502}

\bibitem[{{Krishnan} {et~al.}(2017){Krishnan}, {Hatch}, {Almaini}, {Kocevski},
  {Cooke}, {Hartley}, {Hasinger}, {Maltby}, {Muldrew}, \&
  {Simpson}}]{Krishnan2017a}
{Krishnan}, C., {Hatch}, N.~A., {Almaini}, O., {et~al.} 2017, \mnras, 470,
  2170, \dodoi{10.1093/mnras/stx1315}

\bibitem[{{Lambert} {et~al.}(2024){Lambert}, {Assef}, {Mazzucchelli},
  {Ba{\~n}ados}, {Aravena}, {Barrientos}, {Gonz{\'a}lez-L{\'o}pez}, {Hu},
  {Infante}, {Malhotra}, {Moya-Sierralta}, {Rhoads}, {Valdes}, {Wang}, {Wold},
  \& {Zheng}}]{Lambert2024a}
{Lambert}, T.~S., {Assef}, R.~J., {Mazzucchelli}, C., {et~al.} 2024, \aap, 689,
  A331, \dodoi{10.1051/0004-6361/202449566}

\bibitem[{{Larson} {et~al.}(2023){Larson}, {Finkelstein}, {Kocevski},
  {Hutchison}, {Trump}, {Arrabal Haro}, {Bromm}, {Cleri}, {Dickinson},
  {Fujimoto}, {Kartaltepe}, {Koekemoer}, {Papovich}, {Pirzkal}, {Tacchella},
  {Zavala}, {Bagley}, {Behroozi}, {Champagne}, {Cole}, {Jung}, {Morales},
  {Yang}, {Zhang}, {Zitrin}, {Amor{\'\i}n}, {Burgarella}, {Casey}, {Ch{\'a}vez
  Ortiz}, {Cox}, {Chworowsky}, {Fontana}, {Gawiser}, {Grazian}, {Grogin},
  {Harish}, {Hathi}, {Hirschmann}, {Holwerda}, {Juneau}, {Leung}, {Lucas},
  {McGrath}, {P{\'e}rez-Gonz{\'a}lez}, {Rigby}, {Seill{\'e}}, {Simons}, {de La
  Vega}, {Weiner}, {Wilkins}, {Yung}, \& {Ceers Team}}]{Larson2023a}
{Larson}, R.~L., {Finkelstein}, S.~L., {Kocevski}, D.~D., {et~al.} 2023, \apjl,
  953, L29, \dodoi{10.3847/2041-8213/ace619}

\bibitem[{{Lemaux} {et~al.}(2022){Lemaux}, {Cucciati}, {Le F{\`e}vre},
  {Zamorani}, {Lubin}, {Hathi}, {Ilbert}, {Pelliccia}, {Amor{\'\i}n},
  {Bardelli}, {Cassata}, {Gal}, {Garilli}, {Guaita}, {Giavalisco}, {Hung},
  {Koekemoer}, {Maccagni}, {Pentericci}, {Ribeiro}, {Schaerer}, {Shah}, {Shen},
  {Staab}, {Talia}, {Thomas}, {Tomczak}, {Tresse}, {Vanzella}, {Vergani}, \&
  {Zucca}}]{Lemaux2022a}
{Lemaux}, B.~C., {Cucciati}, O., {Le F{\`e}vre}, O., {et~al.} 2022, \aap, 662,
  A33, \dodoi{10.1051/0004-6361/202039346}

\bibitem[{{Long} {et~al.}(2020){Long}, {Cooray}, {Ma}, {Casey}, {Wardlow},
  {Nayyeri}, {Ivison}, {Farrah}, \& {Dannerbauer}}]{Long2020a}
{Long}, A.~S., {Cooray}, A., {Ma}, J., {et~al.} 2020, \apj, 898, 133,
  \dodoi{10.3847/1538-4357/ab9d1f}

\bibitem[{{Ma} {et~al.}(2016){Ma}, {Hopkins}, {Faucher-Gigu{\`e}re}, {Zolman},
  {Muratov}, {Kere{\v{s}}}, \& {Quataert}}]{Ma2016a}
{Ma}, X., {Hopkins}, P.~F., {Faucher-Gigu{\`e}re}, C.-A., {et~al.} 2016,
  \mnras, 456, 2140, \dodoi{10.1093/mnras/stv2659}

\bibitem[{{Matthee} {et~al.}(2023){Matthee}, {Mackenzie}, {Simcoe}, {Kashino},
  {Lilly}, {Bordoloi}, \& {Eilers}}]{Matthee2023a}
{Matthee}, J., {Mackenzie}, R., {Simcoe}, R.~A., {et~al.} 2023, \apj, 950, 67,
  \dodoi{10.3847/1538-4357/acc846}

\bibitem[{{Matthee} {et~al.}(2024){Matthee}, {Naidu}, {Brammer}, {Chisholm},
  {Eilers}, {Goulding}, {Greene}, {Kashino}, {Labbe}, {Lilly}, {Mackenzie},
  {Oesch}, {Weibel}, {Wuyts}, {Xiao}, {Bordoloi}, {Bouwens}, {van Dokkum},
  {Illingworth}, {Kramarenko}, {Maseda}, {Mason}, {Meyer}, {Nelson}, {Reddy},
  {Shivaei}, {Simcoe}, \& {Yue}}]{Matthee2023b}
{Matthee}, J., {Naidu}, R.~P., {Brammer}, G., {et~al.} 2024, \apj, 963, 129,
  \dodoi{10.3847/1538-4357/ad2345}

\bibitem[{{Mazzucchelli} {et~al.}(2017){Mazzucchelli}, {Ba{\~n}ados},
  {Venemans}, {Decarli}, {Farina}, {Walter}, {Eilers}, {Rix}, {Simcoe},
  {Stern}, {Fan}, {Schlafly}, {De Rosa}, {Hennawi}, {Chambers}, {Greiner},
  {Burgett}, {Draper}, {Kaiser}, {Kudritzki}, {Magnier}, {Metcalfe}, {Waters},
  \& {Wainscoat}}]{Mazzucchelli2017b}
{Mazzucchelli}, C., {Ba{\~n}ados}, E., {Venemans}, B.~P., {et~al.} 2017, \apj,
  849, 91, \dodoi{10.3847/1538-4357/aa9185}

\bibitem[{{Meyer} {et~al.}(2022){Meyer}, {Decarli}, {Walter}, {Li}, {Wang},
  {Mazzucchelli}, {Ba{\~n}ados}, {Farina}, \& {Venemans}}]{Meyer2022a}
{Meyer}, R.~A., {Decarli}, R., {Walter}, F., {et~al.} 2022, \apj, 927, 141,
  \dodoi{10.3847/1538-4357/ac4f67}

\bibitem[{{Morishita} {et~al.}(2023){Morishita}, {Roberts-Borsani}, {Treu},
  {Brammer}, {Mason}, {Trenti}, {Vulcani}, {Wang}, {Acebron}, {Bah{\'e}},
  {Bergamini}, {Boyett}, {Bradac}, {Calabr{\`o}}, {Castellano}, {Chen}, {De
  Lucia}, {Filippenko}, {Fontana}, {Glazebrook}, {Grillo}, {Henry}, {Jones},
  {Kelly}, {Koekemoer}, {Leethochawalit}, {Lu}, {Marchesini}, {Mascia},
  {Mercurio}, {Merlin}, {Metha}, {Nanayakkara}, {Nonino}, {Paris},
  {Pentericci}, {Rosati}, {Santini}, {Strait}, {Vanzella}, {Windhorst}, \&
  {Xie}}]{Morishita2023a}
{Morishita}, T., {Roberts-Borsani}, G., {Treu}, T., {et~al.} 2023, \apjl, 947,
  L24, \dodoi{10.3847/2041-8213/acb99e}

\bibitem[{{Ota} {et~al.}(2018){Ota}, {Venemans}, {Taniguchi}, {Kashikawa},
  {Nakata}, {Harikane}, {Ba{\~n}ados}, {Overzier}, {Riechers}, {Walter},
  {Toshikawa}, {Shibuya}, \& {Jiang}}]{Ota2018a}
{Ota}, K., {Venemans}, B.~P., {Taniguchi}, Y., {et~al.} 2018, \apj, 856, 109,
  \dodoi{10.3847/1538-4357/aab35b}

\bibitem[{{P{\'e}rez-Gonz{\'a}lez} {et~al.}(2023){P{\'e}rez-Gonz{\'a}lez},
  {Barro}, {Annunziatella}, {Costantin}, {Garc{\'\i}a-Argum{\'a}nez},
  {McGrath}, {M{\'e}rida}, {Zavala}, {Arrabal Haro}, {Bagley}, {Backhaus},
  {Behroozi}, {Bell}, {Bisigello}, {Buat}, {Calabr{\`o}}, {Casey}, {Cleri},
  {Coogan}, {Cooper}, {Cooray}, {Dekel}, {Dickinson}, {Elbaz}, {Ferguson},
  {Finkelstein}, {Fontana}, {Franco}, {Gardner}, {Giavalisco},
  {G{\'o}mez-Guijarro}, {Grazian}, {Grogin}, {Guo}, {Huertas-Company}, {Jogee},
  {Kartaltepe}, {Kewley}, {Kirkpatrick}, {Kocevski}, {Koekemoer}, {Long},
  {Lotz}, {Lucas}, {Papovich}, {Pirzkal}, {Ravindranath}, {Somerville},
  {Tacchella}, {Trump}, {Wang}, {Wilkins}, {Wuyts}, {Yang}, \&
  {Yung}}]{PerezGonzalez2023a}
{P{\'e}rez-Gonz{\'a}lez}, P.~G., {Barro}, G., {Annunziatella}, M., {et~al.}
  2023, \apjl, 946, L16, \dodoi{10.3847/2041-8213/acb3a5}

\bibitem[{{P{\'e}rez-Gonz{\'a}lez} {et~al.}(2024){P{\'e}rez-Gonz{\'a}lez},
  {Barro}, {Rieke}, {Lyu}, {Rieke}, {Alberts}, {Williams}, {Hainline}, {Sun},
  {Pusk{\'a}s}, {Annunziatella}, {Baker}, {Bunker}, {Egami}, {Ji}, {Johnson},
  {Robertson}, {Rodr{\'\i}guez Del Pino}, {Rujopakarn}, {Shivaei}, {Tacchella},
  {Willmer}, \& {Willott}}]{PerezGonzalez2024a}
{P{\'e}rez-Gonz{\'a}lez}, P.~G., {Barro}, G., {Rieke}, G.~H., {et~al.} 2024,
  \apj, 968, 4, \dodoi{10.3847/1538-4357/ad38bb}

\bibitem[{{P{\'e}rez-Mart{\'\i}nez}
  {et~al.}(2023{\natexlab{a}}){P{\'e}rez-Mart{\'\i}nez}, {Dannerbauer},
  {Kodama}, {Koyama}, {Shimakawa}, {Suzuki}, {Calvi}, {Chen}, {Daikuhara},
  {Hatch}, {Laza-Ramos}, {Sobral}, {Stott}, \& {Tanaka}}]{PerezMartinez2022a}
{P{\'e}rez-Mart{\'\i}nez}, J.~M., {Dannerbauer}, H., {Kodama}, T., {et~al.}
  2023{\natexlab{a}}, \mnras, 518, 1707, \dodoi{10.1093/mnras/stac2784}

\bibitem[{{P{\'e}rez-Mart{\'\i}nez}
  {et~al.}(2023{\natexlab{b}}){P{\'e}rez-Mart{\'\i}nez}, {Dannerbauer},
  {Kodama}, {Koyama}, {Shimakawa}, {Suzuki}, {Calvi}, {Chen}, {Daikuhara},
  {Hatch}, {Laza-Ramos}, {Sobral}, {Stott}, \& {Tanaka}}]{PerezMartinez2023a}
---. 2023{\natexlab{b}}, \mnras, 518, 1707, \dodoi{10.1093/mnras/stac2784}

\bibitem[{{Remus} {et~al.}(2023){Remus}, {Dolag}, \&
  {Dannerbauer}}]{Remus2020a}
{Remus}, R.-S., {Dolag}, K., \& {Dannerbauer}, H. 2023, \apj, 950, 191,
  \dodoi{10.3847/1538-4357/accb91}

\bibitem[{{Rennehan}(2024)}]{Rennehan2024a}
{Rennehan}, D. 2024, \apj, 975, 114, \dodoi{10.3847/1538-4357/ad793d}

\bibitem[{{Sarrouh} {et~al.}(2024){Sarrouh}, {Muzzin}, {Iyer}, {Mowla},
  {Withers}, {Martis}, {Abraham}, {Asada}, {Brada{\v{c}}}, {Brammer},
  {Desprez}, {Estrada-Carpenter}, {Matharu}, {Noirot}, {Sawicki}, {Strait},
  {Willott}, \& {Zabl}}]{Sarrouh2024a}
{Sarrouh}, G. T.~E., {Muzzin}, A., {Iyer}, K.~G., {et~al.} 2024, \apjl, 967,
  L17, \dodoi{10.3847/2041-8213/ad43e8}

\bibitem[{{Schlafly} \& {Finkbeiner}(2011)}]{Schlafly2011a}
{Schlafly}, E.~F., \& {Finkbeiner}, D.~P. 2011, \apj, 737, 103,
  \dodoi{10.1088/0004-637X/737/2/103}

\bibitem[{{Schlegel} {et~al.}(1998){Schlegel}, {Finkbeiner}, \&
  {Davis}}]{Schlegel1998a}
{Schlegel}, D.~J., {Finkbeiner}, D.~P., \& {Davis}, M. 1998, \apj, 500, 525,
  \dodoi{10.1086/305772}

\bibitem[{{Scholtz} {et~al.}(2023){Scholtz}, {Maiolino}, {D'Eugenio},
  {Curtis-Lake}, {Carniani}, {Charlot}, {Curti}, {Silcock}, {Arribas}, {Baker},
  {Bhatawdekar}, {Boyett}, {Bunker}, {Chevallard}, {Circosta}, {Eisenstein},
  {Hainline}, {Hausen}, {Ji}, {Ji}, {Johnson}, {Kumari}, {Looser}, {Lyu},
  {Maseda}, {Parlanti}, {Perna}, {Rieke}, {Robertson}, {Rodr{\'\i}guez Del
  Pino}, {Sun}, {Tacchella}, {{\"U}bler}, {Venturi}, {Williams}, {Willmer},
  {Willott}, \& {Witstok}}]{Scholtz2023a}
{Scholtz}, J., {Maiolino}, R., {D'Eugenio}, F., {et~al.} 2023, arXiv e-prints,
  arXiv:2311.18731, \dodoi{10.48550/arXiv.2311.18731}

\bibitem[{{Shapley} {et~al.}(2023){Shapley}, {Reddy}, {Sanders}, {Topping}, \&
  {Brammer}}]{Shapley2023a}
{Shapley}, A.~E., {Reddy}, N.~A., {Sanders}, R.~L., {Topping}, M.~W., \&
  {Brammer}, G.~B. 2023, \apjl, 950, L1, \dodoi{10.3847/2041-8213/acd939}

\bibitem[{{Shivaei} {et~al.}(2024){Shivaei}, {Alberts}, {Florian}, {Rieke},
  {Wuyts}, {Bodansky}, {Bunker}, {Cameron}, {Curti}, {D'Eugenio},
  {Dudzevi{\v{c}}i{\={u}}t{\.{e}}}, {Ji}, {Johnson}, {Kramarenko}, {Lyu},
  {Matthee}, {Morrison}, {Naidu}, {P{\'e}rez-Gonz{\'a}lez}, {Reddy},
  {Robertson}, {Sun}, {Tacchella}, {Whitaker}, {Williams}, {Willmer},
  {Witstok}, {Xiao}, \& {Zhu}}]{Shivaei2024a}
{Shivaei}, I., {Alberts}, S., {Florian}, M., {et~al.} 2024, \aap, 690, A89,
  \dodoi{10.1051/0004-6361/202449579}

\bibitem[{{Sun} {et~al.}(2023){Sun}, {Helton}, {Egami}, {Hainline}, {Rieke},
  {Rieke}, {Willmer}, \& {Jades Collaboration}}]{Sun2023b}
{Sun}, F., {Helton}, J., {Egami}, E., {et~al.} 2023, in American Astronomical
  Society Meeting Abstracts, Vol. 242, American Astronomical Society Meeting
  Abstracts, 206.02

\bibitem[{{Tang} {et~al.}(2019){Tang}, {Stark}, {Chevallard}, \&
  {Charlot}}]{Tang2019a}
{Tang}, M., {Stark}, D.~P., {Chevallard}, J., \& {Charlot}, S. 2019, \mnras,
  489, 2572, \dodoi{10.1093/mnras/stz2236}

\bibitem[{{Topping} {et~al.}(2022){Topping}, {Stark}, {Endsley}, {Bouwens},
  {Schouws}, {Smit}, {Stefanon}, {Inami}, {Bowler}, {Oesch}, {Gonzalez},
  {Dayal}, {da Cunha}, {Algera}, {van der Werf}, {Pallottini}, {Barrufet},
  {Schneider}, {De Looze}, {Sommovigo}, {Whitler}, {Graziani}, {Fudamoto}, \&
  {Ferrara}}]{Topping2022a}
{Topping}, M.~W., {Stark}, D.~P., {Endsley}, R., {et~al.} 2022, \mnras, 516,
  975, \dodoi{10.1093/mnras/stac2291}

\bibitem[{{Venemans} {et~al.}(2019){Venemans}, {Neeleman}, {Walter}, {Novak},
  {Decarli}, {Hennawi}, \& {Rix}}]{Venemans2019a}
{Venemans}, B.~P., {Neeleman}, M., {Walter}, F., {et~al.} 2019, \apjl, 874,
  L30, \dodoi{10.3847/2041-8213/ab11cc}

\bibitem[{{Venemans} {et~al.}(2016){Venemans}, {Walter}, {Zschaechner},
  {Decarli}, {De Rosa}, {Findlay}, {McMahon}, \& {Sutherland}}]{Venemans2016a}
{Venemans}, B.~P., {Walter}, F., {Zschaechner}, L., {et~al.} 2016, \apj, 816,
  37, \dodoi{10.3847/0004-637X/816/1/37}

\bibitem[{{Venemans} {et~al.}(2013){Venemans}, {Findlay}, {Sutherland}, {De
  Rosa}, {McMahon}, {Simcoe}, {Gonz{\'a}lez-Solares}, {Kuijken}, \&
  {Lewis}}]{Venemans2013a}
{Venemans}, B.~P., {Findlay}, J.~R., {Sutherland}, W.~J., {et~al.} 2013, \apj,
  779, 24, \dodoi{10.1088/0004-637X/779/1/24}

\bibitem[{{Villa-V{\'e}lez} {et~al.}(2021){Villa-V{\'e}lez}, {Buat},
  {Theul{\'e}}, {Boquien}, \& {Burgarella}}]{VillaVelez2021a}
{Villa-V{\'e}lez}, J.~A., {Buat}, V., {Theul{\'e}}, P., {Boquien}, M., \&
  {Burgarella}, D. 2021, \aap, 654, A153, \dodoi{10.1051/0004-6361/202140890}

\bibitem[{{Vito} {et~al.}(2020){Vito}, {Brandt}, {Lehmer}, {Vignali}, {Zou},
  {Bauer}, {Bremer}, {Gilli}, {Ivison}, \& {Spingola}}]{Vito2020a}
{Vito}, F., {Brandt}, W.~N., {Lehmer}, B.~D., {et~al.} 2020, \aap, 642, A149,
  \dodoi{10.1051/0004-6361/202038848}

\bibitem[{{Vito} {et~al.}(2024){Vito}, {Brandt}, {Comastri}, {Gilli}, {Ivison},
  {Lanzuisi}, {Lehmer}, {Lopez}, {Tozzi}, \& {Vignali}}]{Vito2024a}
{Vito}, F., {Brandt}, W.~N., {Comastri}, A., {et~al.} 2024, \aap, 689, A130,
  \dodoi{10.1051/0004-6361/202450225}

\bibitem[{{Wang} {et~al.}(2023){Wang}, {Yang}, {Hennawi}, {Fan}, {Sun},
  {Champagne}, {Costa}, {Habouzit}, {Endsley}, {Li}, {Lin}, {Meyer},
  {Schindler}, {Wu}, {Ba{\~n}ados}, {Barth}, {Bhowmick}, {Bieri}, {Blecha},
  {Bosman}, {Cai}, {Colina}, {Connor}, {Davies}, {Decarli}, {De Rosa}, {Drake},
  {Egami}, {Eilers}, {Evans}, {Farina}, {Haiman}, {Jiang}, {Jin}, {Jun},
  {Kakiichi}, {Khusanova}, {Kulkarni}, {Li}, {Liu}, {Loiacono}, {Lupi},
  {Mazzucchelli}, {Onoue}, {Pudoka}, {Rojas-Ruiz}, {Shen}, {Strauss}, {Tee},
  {Trakhtenbrot}, {Trebitsch}, {Venemans}, {Volonteri}, {Walter}, {Xie}, {Yue},
  {Zhang}, {Zhang}, \& {Zou}}]{Wang2023a}
{Wang}, F., {Yang}, J., {Hennawi}, J.~F., {et~al.} 2023, \apjl, 951, L4,
  \dodoi{10.3847/2041-8213/accd6f}

\bibitem[{{Whitler} {et~al.}(2023){Whitler}, {Stark}, {Endsley}, {Leja},
  {Charlot}, \& {Chevallard}}]{Whitler2023a}
{Whitler}, L., {Stark}, D.~P., {Endsley}, R., {et~al.} 2023, \mnras, 519, 5859,
  \dodoi{10.1093/mnras/stad004}

\bibitem[{{Wilkins} {et~al.}(2023){Wilkins}, {Lovell}, {Vijayan}, {Irodotou},
  {Adams}, {Roper}, {Caruana}, {Matthee}, {Seeyave}, {Conselice},
  {P{\'e}rez-Gonz{\'a}lez}, {Turner}, {Donnellan}, {Verma}, \&
  {Trussler}}]{Wilkins2023a}
{Wilkins}, S.~M., {Lovell}, C.~C., {Vijayan}, A.~P., {et~al.} 2023, \mnras,
  522, 4014, \dodoi{10.1093/mnras/stad1126}

\bibitem[{{Yang} {et~al.}(2023){Yang}, {Wang}, {Fan}, {Hennawi}, {Barth},
  {Ba{\~n}ados}, {Sun}, {Liu}, {Cai}, {Jiang}, {Li}, {Onoue}, {Schindler},
  {Shen}, {Wu}, {Bhowmick}, {Bieri}, {Blecha}, {Bosman}, {Champagne}, {Colina},
  {Connor}, {Costa}, {Davies}, {Decarli}, {De Rosa}, {Drake}, {Egami},
  {Eilers}, {Evans}, {Farina}, {Habouzit}, {Haiman}, {Jin}, {Jun}, {Kakiichi},
  {Khusanova}, {Kulkarni}, {Loiacono}, {Lupi}, {Mazzucchelli}, {Pan},
  {Rojas-Ruiz}, {Strauss}, {Tee}, {Trakhtenbrot}, {Trebitsch}, {Venemans},
  {Vestergaard}, {Volonteri}, {Walter}, {Xie}, {Yue}, {Zhang}, {Zhang}, \&
  {Zou}}]{Yang2023a}
{Yang}, J., {Wang}, F., {Fan}, X., {et~al.} 2023, \apjl, 951, L5,
  \dodoi{10.3847/2041-8213/acc9c8}

\bibitem[{{Zhang} {et~al.}(2024){Zhang}, {Behroozi}, {Volonteri}, {Silk},
  {Fan}, {Aird}, {Yang}, {Wang}, {Tee}, \& {Hopkins}}]{Zhang2023d}
{Zhang}, H., {Behroozi}, P., {Volonteri}, M., {et~al.} 2024, \mnras, 531, 4974,
  \dodoi{10.1093/mnras/stae1447}

\end{thebibliography}
\bibliographystyle{aasjournal}



\suppressAffiliationsfalse
\allauthors

\end{document}